\PassOptionsToPackage{unicode}{hyperref}
\PassOptionsToPackage{hyphens}{url}
\PassOptionsToPackage{dvipsnames,svgnames,x11names}{xcolor}
\documentclass[
  9pt,
]{article}
\usepackage[a4paper, total={7in, 10in}]{geometry}

\usepackage{amsmath,amssymb}
\usepackage{lmodern}
\usepackage{iftex}
\ifPDFTeX
  \usepackage[T1]{fontenc}
  \usepackage[utf8]{inputenc}
  \usepackage{textcomp} 
\else 
  \usepackage{unicode-math}
  \defaultfontfeatures{Scale=MatchLowercase}
  \defaultfontfeatures[\rmfamily]{Ligatures=TeX,Scale=1}
\fi
\IfFileExists{upquote.sty}{\usepackage{upquote}}{}
\IfFileExists{microtype.sty}{
  \usepackage[]{microtype}
  \UseMicrotypeSet[protrusion]{basicmath} 
}{}
\makeatletter
\@ifundefined{KOMAClassName}{
  \IfFileExists{parskip.sty}{%
    \usepackage{parskip}
  }{
    \setlength{\parindent}{0pt}
    \setlength{\parskip}{6pt plus 2pt minus 1pt}}
}{
  \KOMAoptions{parskip=half}}
\makeatother
\usepackage{xcolor}
\ifx\paragraph\undefined\else
  \let\oldparagraph\paragraph
  \renewcommand{\paragraph}[1]{\oldparagraph{#1}\mbox{}}
\fi
\ifx\subparagraph\undefined\else
  \let\oldsubparagraph\subparagraph
  \renewcommand{\subparagraph}[1]{\oldsubparagraph{#1}\mbox{}}
\fi

\usepackage{color}
\usepackage{fancyvrb}

\DefineVerbatimEnvironment{Highlighting}{Verbatim}{commandchars=\\\{\}}
\usepackage{framed}
\definecolor{shadecolor}{RGB}{241,243,245}
\newenvironment{Shaded}{\begin{snugshade}}{\end{snugshade}}

\newcommand{\AttributeTok}[1]{\textcolor[rgb]{0.40,0.45,0.13}{#1}}

\newcommand{\CommentTok}[1]{\textcolor[rgb]{0.37,0.37,0.37}{#1}}

\newcommand{\ConstantTok}[1]{\textcolor[rgb]{0.56,0.35,0.01}{#1}}
\newcommand{\ControlFlowTok}[1]{\textcolor[rgb]{0.00,0.23,0.31}{#1}}

\newcommand{\DecValTok}[1]{\textcolor[rgb]{0.68,0.00,0.00}{#1}}

\newcommand{\FloatTok}[1]{\textcolor[rgb]{0.68,0.00,0.00}{#1}}
\newcommand{\FunctionTok}[1]{\textcolor[rgb]{0.28,0.35,0.67}{#1}}

\newcommand{\NormalTok}[1]{\textcolor[rgb]{0.00,0.23,0.31}{#1}}

\newcommand{\OtherTok}[1]{\textcolor[rgb]{0.00,0.23,0.31}{#1}}

\newcommand{\SpecialCharTok}[1]{\textcolor[rgb]{0.37,0.37,0.37}{#1}}

\newcommand{\StringTok}[1]{\textcolor[rgb]{0.13,0.47,0.30}{#1}}

\providecommand{\tightlist}{%
  \setlength{\itemsep}{0pt}\setlength{\parskip}{0pt}}\usepackage{longtable,booktabs,array}
\usepackage{calc} 
\usepackage{etoolbox}
\makeatletter
\patchcmd\longtable{\par}{\if@noskipsec\mbox{}\fi\par}{}{}
\makeatother
\IfFileExists{footnotehyper.sty}{\usepackage{footnotehyper}}{\usepackage{footnote}}
\makesavenoteenv{longtable}
\usepackage{graphicx}
\makeatletter
\def\maxwidth{\ifdim\Gin@nat@width>\linewidth\linewidth\else\Gin@nat@width\fi}
\def\maxheight{\ifdim\Gin@nat@height>\textheight\textheight\else\Gin@nat@height\fi}
\makeatother
\setkeys{Gin}{width=\maxwidth,height=\maxheight,keepaspectratio}
\makeatletter
\def\fps@figure{htbp}
\makeatother
\newlength{\cslhangindent}
\newlength{\csllabelwidth}
\newlength{\cslentryspacingunit} 
\newenvironment{CSLReferences}[2] 
 {
  \setlength{\parindent}{0pt}
  \ifodd #1
  \let\oldpar\par
  \def\par{\hangindent=\cslhangindent\oldpar}
  \fi
  \setlength{\parskip}{#2\cslentryspacingunit}
 }%
 {}
\usepackage{calc}

\newcommand{\CSLLeftMargin}[1]{\parbox[t]{\csllabelwidth}{#1}}
\newcommand{\CSLRightInline}[1]{\parbox[t]{\linewidth - \csllabelwidth}{#1}\break}

\usepackage[noblocks]{authblk}

\makeatletter
\@ifpackageloaded{caption}{}{\usepackage{caption}}
\AtBeginDocument{%
\ifdefined\contentsname
  \renewcommand*\contentsname{Table of contents}
\else
  \newcommand\contentsname{Table of contents}
\fi
\ifdefined\listfigurename
  \renewcommand*\listfigurename{List of Figures}
\else
  \newcommand\listfigurename{List of Figures}
\fi
\ifdefined\listtablename
  \renewcommand*\listtablename{List of Tables}
\else
  \newcommand\listtablename{List of Tables}
\fi
\ifdefined\figurename
  \renewcommand*\figurename{Figure}
\else
  \newcommand\figurename{Figure}
\fi
\ifdefined\tablename
  \renewcommand*\tablename{Table}
\else
  \newcommand\tablename{Table}
\fi
}
\@ifpackageloaded{float}{}{\usepackage{float}}
\floatstyle{ruled}
\@ifundefined{c@chapter}{\newfloat{codelisting}{h}{lop}}{\newfloat{codelisting}{h}{lop}[chapter]}
\floatname{codelisting}{Listing}

\makeatother
\makeatletter
\@ifpackageloaded{caption}{}{\usepackage{caption}}
\@ifpackageloaded{subcaption}{}{\usepackage{subcaption}}
\makeatother
\makeatletter
\@ifpackageloaded{tcolorbox}{}{\usepackage[many]{tcolorbox}}
\makeatother
\makeatletter
\@ifundefined{shadecolor}{\definecolor{shadecolor}{rgb}{.97, .97, .97}}
\makeatother
\ifLuaTeX
  \usepackage{selnolig}  
\fi
\IfFileExists{bookmark.sty}{\usepackage{bookmark}}{\usepackage{hyperref}}
\IfFileExists{xurl.sty}{\usepackage{xurl}}{} 
\hypersetup{
  pdftitle={Temporal network analysis: Introduction, methods and detailed analysis with R},
  pdfkeywords={temporal networks, temporal analysis, social network
analysis, learning analytics},
  colorlinks=true,
  linkcolor={blue},
  filecolor={Maroon},
  citecolor={Blue},
  urlcolor={Blue},
  pdfcreator={LaTeX via pandoc}}

\title{Temporal network analysis: Introduction, methods and detailed tutorial  
with R}

  \author{Mohammed Saqr}
            \affil{%
                  University of Eastern Finland
              }
      
\begin{document}
\maketitle
\begin{abstract}
Learning involves relations, interactions and connections between
learners, teachers and the world at large. Such interactions are
essentially temporal and unfold in time. Yet, researchers have rarely
combined the two aspects (the temporal and relational aspects) in an
analytics framework. Temporal networks allow modeling of the temporal
learning processes i.e., the emergence and flow of activities,
communities, and social processes through fine-grained dynamic analysis.
This can provide insights into phenomena like knowledge co-construction,
information flow, and relationship building. This chapter introduces the
basic concepts of temporal networks, their types and techniques. A
detailed guide of temporal network analysis is introduced in this
chapter, that starts with building the network, visualization,
mathematical analysis on the node and graph level. The analysis is
performed with a real-world dataset. The discussion chapter offers some
extra resources for interested users who want to expand their knowledge
of the technique.
\end{abstract}
\ifdefined\Shaded\renewenvironment{Shaded}{\begin{tcolorbox}[frame hidden, interior hidden, boxrule=0pt, borderline west={3pt}{0pt}{shadecolor}, enhanced, sharp corners, breakable]}{\end{tcolorbox}}\fi

\hypertarget{introduction}{%
\section{Introduction}\label{introduction}}

Learning is social and therefore, involves relations, interactions and
connections between learners, teachers and the world at large. Such
interactions are essentially temporal and unfold in time {[}1{]}. That
is facilitated, curtailed or influenced at different temporal scales
{[}2, 3{]}. Therefore, time has become a quintessential aspect in
several learning theories, frameworks and methodological approaches to
learning {[}3--5{]}. Modeling learning as a temporal and relational
process is, nevertheless, both natural, timely and more tethered to
reality {[}4, 6{]}. Traditionally, relations have been modeled with
Social Network Analysis (SNA) and temporal events have been modeled with
sequence analysis or process mining {[}3, 7{]}.Yet, researchers have
rarely combined the two aspects (the temporal and relational aspects) in
an analytics framework {[}1{]}. Considering how important the timing and
order of the learning process are, it is all-important that our analysis
lens is not time-blind {[}8, 9{]}. Using time-blind methods flattens an
essentially temporal process where the important details of progression
are lost or distorted {[}10, 11{]}. In doing so, we miss the rhythm, the
evolution and devolution of the process, we overlook the regularity and
we may fail to capture the events that matter {[}9--11{]}.

\textbf{Temporal networks}

Recent advances in network analysis have resulted in the emergence of
the new field of temporal network analysis which combines both the
relational and temporal dimensions into a single analytical framework:
temporal networks, also referred to as time-varying networks, dynamic
networks or evolving networks {[}10{]}. Today, temporal networks are
increasingly adopted in several fields to model dynamic phenomena, e.g.,
information exchange, the spread of infections, or the reach of viral
videos on social media {[}12{]}. Whereas temporal networks are concerned
with the modeling of relationships similar to traditional social
networks (i.e., static or aggregate networks), they are conceptually
fundamentally different {[}10, 11, 13{]}. Additionally, temporal
networks are not a simple extension of social networks, nor are they
time-augmented social networks or time-weighted networks. As Holme puts
it, ``temporal-network modeling is far from a straightforward
generalization of static networks---often, it is fundamentally
different'' {[}12{]}.

\begin{figure}

{\centering \includegraphics[width=6in,height=\textheight]{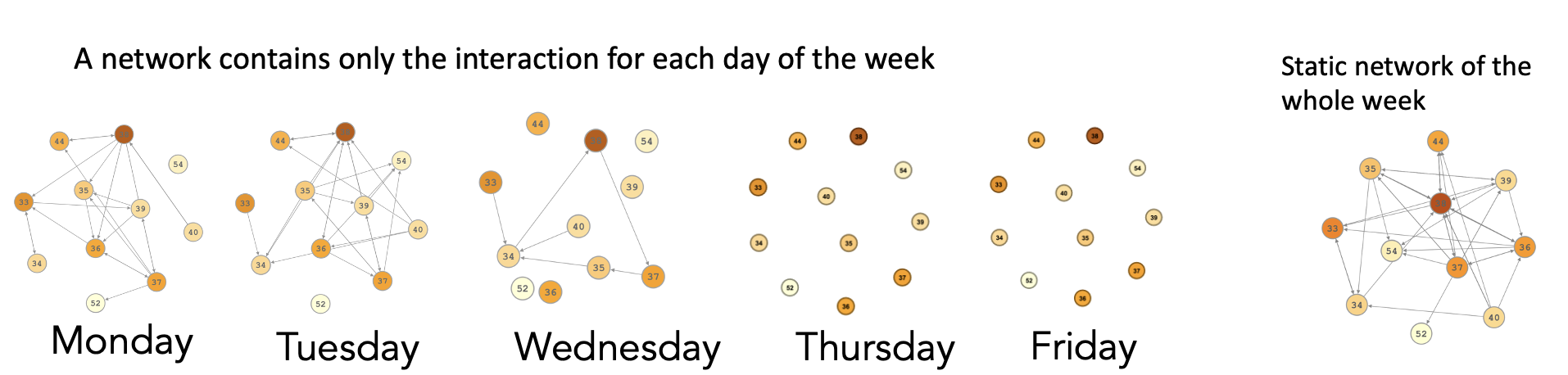}

}

\caption{\label{fig-interactions}Interactions are aggregated per day
showing that some days show an active group e.g., Monday, and other days
have an inactive group e.g., Friday. Meanwhile, the static network on
the right appears dense {[}15{]}}

\end{figure}

Temporal networks differ significantly from SNA, they are based on
different representations of data, have a different mathematical
underpinning, and use distinct visualization methods. In temporal
networks, edges emerge (get activated or born) and dissolve (get
deactivated or die) compared to always present edges in social networks.
Also, in temporal networks, edges are a temporary interaction, contact,
co-presence, or concurrency between two nodes interacting at a specific
time. The fact that static networks represent nodes as connected
together all the time exaggerates connectivity {[}14, 15{]}. For
instance, in Figure~\ref{fig-interactions}, we have five network
visualizations, each network belonging to a weekday. We see that Monday,
Tuesday, and Wednesday networks are relatively connected, whereas
Thursday and Friday networks are disconnected. The corresponding
aggregated or static network on the right is densely connected. The
example in Figure~\ref{fig-interactions} shows clearly how a static
network both conflates connectivity and obfuscates dynamics, you can
read more about this example in {[}15{]}. Another characteristic of
temporal networks is that edges have a starting time point and ending
time point, the end of each edge is understandably later than the start,
i.e., follows the forward-moving direction of time. Therefore, the paths
between edges in the temporal network are unidirectional or
time-restricted {[}10, 11{]}. The next section discusses the differences
between traditional networks and temporal networks in detail.

\hypertarget{the-building-blocks-of-a-temporal-network}{%
\section{The building blocks of a temporal
network}\label{the-building-blocks-of-a-temporal-network}}

\hypertarget{edges}{%
\subsection{Edges}\label{edges}}

In temporal networks, edges are commonly referred to as events, links,
or dynamic edges. Two types of temporal networks are commonly described
based on their edge type {[}12{]}.

\begin{figure}

{\centering \includegraphics[width=4.5in,height=\textheight]{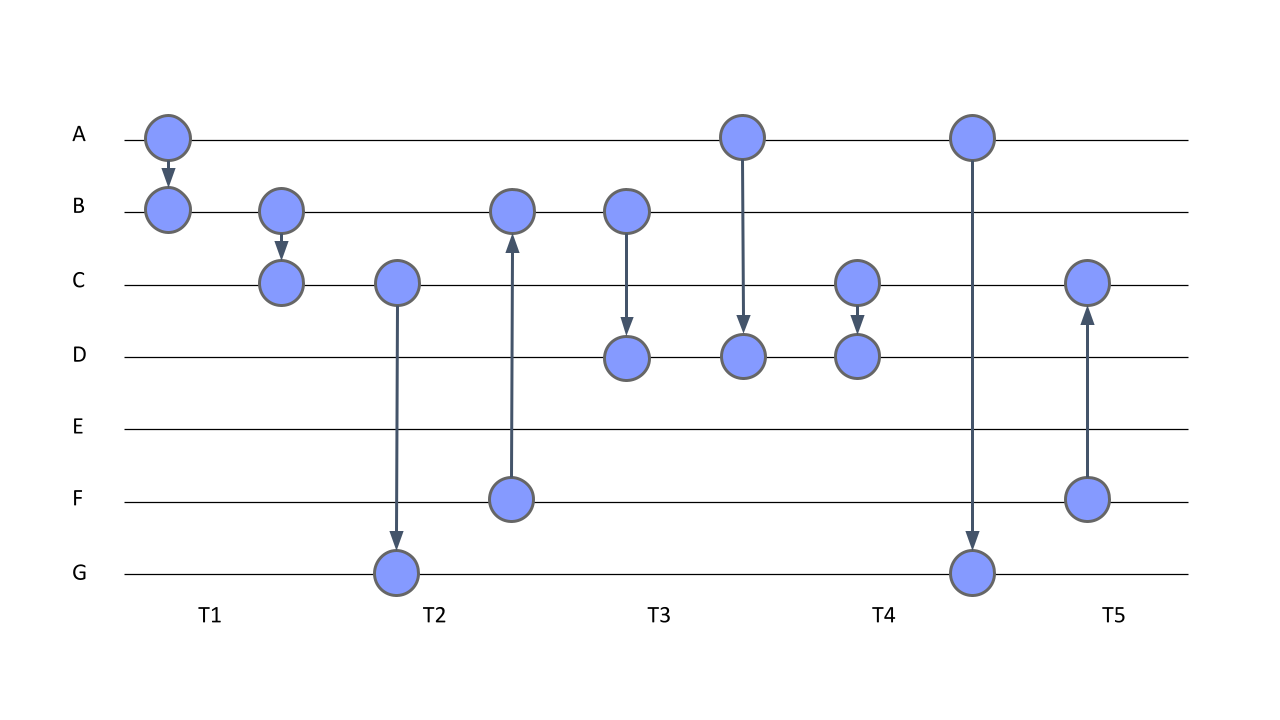}

}

\caption{\label{fig-contact}Example of a contact temporal network. Edges
form momentarily and have no obvious duration.}

\end{figure}

\begin{itemize}
\tightlist
\item
  \textbf{Contact temporal networks:} In contact temporal networks, edge
  duration is very brief, undefined, or negligible. For example, instant
  messages have no obvious duration but have a clear source (sender),
  target (receiver), and timestamp. Figure~\ref{fig-contact} shows a
  contact temporal network where the edges are represented as sequences
  of contacts between nodes with no duration.
\end{itemize}

\begin{figure}

{\centering \includegraphics[width=4.5in,height=\textheight]{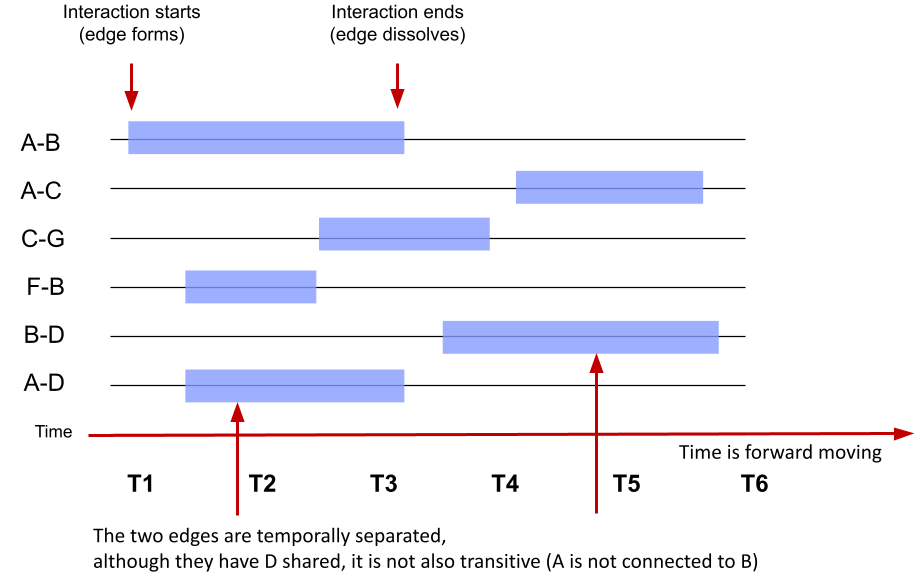}

}

\caption{\label{fig-interval}An example of an interval network. On the
top, we see an example of an edge. In the bottom arrows pointing to two
edges B-D and A-D, we can see that the two edges do not overlap and
therefore, although they share a connection (A), we can't assume that B
is connected to A through D.}

\end{figure}

\begin{itemize}
\tightlist
\item
  \textbf{Interval temporal networks:} In interval temporal networks,
  each interaction has a duration. An example of such a network would be
  a conversation where each of the conversants talks for a certain
  length of time. In the interval temporal network, the duration of
  interactions matters and the modeling thereof helps understand the
  process. In Figure~\ref{fig-interval}, we see an interval temporal
  network where each edge has a clear start and clear end. For example,
  an edge forms between node A and node B at time point 1 and dissolves
  at time point 3, i.e., lasts for two time points.
\end{itemize}

\hypertarget{paths-concurrency-and-reachability}{%
\subsection{Paths, concurrency, and
reachability}\label{paths-concurrency-and-reachability}}

Paths represent the pathways that connect nodes, the identification of
which can help solve essential problems like the shortest path between
two places in a route planning application, e.g., Google maps. In a
dynamic process, the paths represent a time-respecting sequence of edges
i.e., where the timing of each edge follows one another according to
time passage, that is, the timestamps are incrementally increasing
{[}10, 11{]}. For instance, let's assume we have a group of students
interacting about a problem, starting by defining the problem,
argumenting, debating, and finding a solution. The temporal path that
would represent the sequence of interactions among students in this
process will be a
\emph{defining}-\textgreater{}\emph{argumenting}-\textgreater{}\emph{debating}-\textgreater{}\emph{solving}.
We expect that the timestamp of \emph{defining} precedes
\emph{argumenting} and \emph{argumenting} precedes \emph{debating} and
so on. In that way, the path is unidirectional, follows a time-ordered
sequence, and requires that each node is temporally connected, i.e., the
two nodes coexist or interact with each other at the same time {[}11{]}.
Such temporal co-presence is known as concurrent. Concurrency defines
the duration of the nodes where they were co-present together and
therefore can be a measure of the magnitude of contact between the two
nodes. This is particularly important when we are modeling processes
where the length of interactions matter e.g., social influence. A
student is more likely to be influenced by an idea when the student
discusses the idea with another for a longer period of time. Similarly,
self-regulation could be more meaningful when phases are more concurrent
rather than disconnected {[}3{]}.

\hypertarget{literature-review-of-prior-work}{%
\section{Literature review of prior
work}\label{literature-review-of-prior-work}}

Few studies have addressed temporal network analysis. Yet, some examples
exist that may shed light on the novel framework and how it can be
harnessed in education. In a study by Saqr and Nouri {[}15{]}, the
authors investigated how students interact in a problem-based learning
environment using temporal networks. The study estimated temporal
centrality measures, used temporal network visualization, and examined
the predictive power of temporal centrality measures. The study reported
rhythmic changes in centrality measures, network properties as well as
the way students mix online. The study also found that temporal
centrality measures were predictive of students'' performance from as
early as the second day of the course. Models that included temporal
centrality measures have performed consistently better and from as early
as the first week of the course. Another study by {[}9{]} analyzed
students' interactions in an online collaborative environment where
students interacted in Facebook groups. The authors compared centrality
measures from traditional social networks to temporal centrality
measures and found that temporal centralities are more predictive of
performance. Another study from the same group has used chat messages to
study how students interact online and how temporal networks can shed
light on different dynamics of students interacting using Discord
instant messaging platform compared to students interacting using the
forums in Moodle. Temporal networks were more informative in capturing
the differences in dynamics and how such dynamics affected students''
way of communicating {[}16{]}.

\hypertarget{tutorial-building-a-temporal-network}{%
\section{Tutorial: Building a temporal
network}\label{tutorial-building-a-temporal-network}}

The first step is to load the needed packages. Unlike the SNA chapter
{[}17{]} where we relied on the \texttt{igraph} framework, we will rely
on the \texttt{statnet} framework that has a rich repertoire of temporal
network packages. We will use three main packages, namely \texttt{tsna}
(Temporal Social Network Analysis) which provides most functions for
dealing with temporal networks as an extension for the popular
\texttt{sna} package. The package \texttt{networkDynamic} offers several
complementary functions for the network manipulation, whereas the
package \texttt{ndtv} (Network Dynamic Temporal Visualization) offers
several functions for visualizing temporal networks. To learn more about
these packages, please visit their help pages. The next code chunk loads
these packages as well as \texttt{tidyverse} packages to process the
network dataframe {[}20{]}. We also need \texttt{tidyverse} for
manipulating the file and preparing the data.

\begin{Shaded}
\begin{Highlighting}[]
\FunctionTok{library}\NormalTok{(tsna)}
\FunctionTok{library}\NormalTok{(ndtv)}
\FunctionTok{library}\NormalTok{(networkDynamic)}
\FunctionTok{library}\NormalTok{(tidyverse)}
\FunctionTok{library}\NormalTok{(rio)}
\end{Highlighting}
\end{Shaded}

To create a temporal network, we need a timestamped file with
interactions. The essential fields are the \texttt{source},
\texttt{target} and \texttt{time}, and perhaps also some information
about the interactions or the nodes (but these constitute extra
information that is good to have). A temporal network is created by
combining a base static network (that has the network base information)
and a dynamic network with time information. As such, we need to prepare
the MOOC dataset described in detail here {[}21{]} and prepare it for
creating a static network that will serve as a base network.

The next code chunk loads the dataset files (edges and nodes data) from
the MOOCs dataset. Some cleaning of the data is necessary. \scriptsize

\begin{Shaded}
\begin{Highlighting}[]
\NormalTok{net\_edges }\OtherTok{\textless{}{-}} \FunctionTok{import}\NormalTok{(}\StringTok{"https://raw.githubusercontent.com/sonsoleslp/labook{-}data/main/6\_snaMOOC/DLT1\%20Edgelist.csv"}\NormalTok{)}
\NormalTok{net\_nodes }\OtherTok{\textless{}{-}} \FunctionTok{import}\NormalTok{(}\StringTok{"https://raw.githubusercontent.com/sonsoleslp/labook{-}data/main/6\_snaMOOC/DLT1\%20Nodes.csv"}\NormalTok{)}
\end{Highlighting}
\end{Shaded}

\normalsize

First, we have to clean the column names from extra spaces using the
function \texttt{clean\_names} from the \texttt{janitor} package. Next,
we have to remove loops, or instances where the source and target of the
interaction are the same since it makes little sense that a person
responds to oneself in a temporal network (this is not essential).
Third, we need to create a dataframe where we replace duplicate edges
with a weight equal to the frequency of repeated interactions, we will
need this file for the creation of the base network (see later). Fourth,
we recode the expertise level in the nodes file to meaningful codes
(from its original numerical coding as 1,2,3) so that we can use them
later in the analysis. The fifth step is to convert the timestamp to
sequential days starting from the first day of the course; this makes
sense for easy interpretation. Also, time works better in networkDynamic
when it is numeric. The final step is to remove discussions where there
are no replies. This cleaning is necessary since we have a dataset that
was not essentially prepared for temporal networks.

\begin{Shaded}
\begin{Highlighting}[]
\NormalTok{net\_edges }\OtherTok{\textless{}{-}}\NormalTok{ net\_edges }\SpecialCharTok{\%\textgreater{}\%}\NormalTok{ janitor}\SpecialCharTok{::}\FunctionTok{clean\_names}\NormalTok{() }\CommentTok{\#1 cleaning column names}
\NormalTok{net\_edges\_NL }\OtherTok{\textless{}{-}}\NormalTok{ net\_edges }\SpecialCharTok{\%\textgreater{}\%} \FunctionTok{filter}\NormalTok{(sender }\SpecialCharTok{!=}\NormalTok{ receiver) }\CommentTok{\#2 removing loops}

\CommentTok{\# Removing duplicates and replacing them with weight}
\NormalTok{net\_edges\_NLW }\OtherTok{\textless{}{-}}\NormalTok{ net\_edges\_NL }\SpecialCharTok{\%\textgreater{}\%} \FunctionTok{group\_by}\NormalTok{(sender, receiver) }\SpecialCharTok{\%\textgreater{}\%} \FunctionTok{tally}\NormalTok{(}\AttributeTok{name=}\StringTok{"weight"}\NormalTok{) }\CommentTok{\#3}

\CommentTok{\# Recoding expertise}
\NormalTok{net\_nodes }\OtherTok{\textless{}{-}}\NormalTok{ net\_nodes }\SpecialCharTok{\%\textgreater{}\%} 
  \FunctionTok{mutate}\NormalTok{(}\AttributeTok{expert\_level=}\FunctionTok{case\_match}\NormalTok{(experience, }\CommentTok{\#4}
        \DecValTok{1} \SpecialCharTok{\textasciitilde{}}\StringTok{"Expert"}\NormalTok{,}
        \DecValTok{2} \SpecialCharTok{\textasciitilde{}} \StringTok{"Student"}\NormalTok{,}
        \DecValTok{3} \SpecialCharTok{\textasciitilde{}} \StringTok{"Teacher"}\NormalTok{))}

\CommentTok{\# A function to create serial days}
\NormalTok{dayizer }\OtherTok{=} \ControlFlowTok{function}\NormalTok{(my\_date) \{}
\NormalTok{  numeric\_date }\OtherTok{=}\NormalTok{ lubridate}\SpecialCharTok{::}\FunctionTok{parse\_date\_time}\NormalTok{(my\_date, }\StringTok{"mdy HM"}\NormalTok{)}
\NormalTok{  Min\_time }\OtherTok{=} \FunctionTok{min}\NormalTok{(numeric\_date)}
\NormalTok{  my\_date }\OtherTok{=}\NormalTok{ (numeric\_date }\SpecialCharTok{{-}}\NormalTok{ Min\_time) }\SpecialCharTok{/}\NormalTok{ (}\DecValTok{24}\SpecialCharTok{*}\DecValTok{60}\SpecialCharTok{*}\DecValTok{60}\NormalTok{)}
\NormalTok{  my\_date }\OtherTok{=} \FunctionTok{round}\NormalTok{(my\_date,}\DecValTok{2}\NormalTok{)}
  \FunctionTok{return}\NormalTok{(}\FunctionTok{as.numeric}\NormalTok{(my\_date))}
\NormalTok{\}}

\NormalTok{net\_edges\_NL}\SpecialCharTok{\$}\NormalTok{new\_date }\OtherTok{=} \FunctionTok{dayizer}\NormalTok{(net\_edges\_NL}\SpecialCharTok{\$}\NormalTok{timestamp) }\CommentTok{\#5}

\CommentTok{\# Remove dicussions with no interactions}
\NormalTok{net\_edges\_NL }\OtherTok{\textless{}{-}}\NormalTok{ net\_edges\_NL }\SpecialCharTok{\%\textgreater{}\%} \FunctionTok{group\_by}\NormalTok{ (discussion\_title) }\SpecialCharTok{\%\textgreater{}\%} \FunctionTok{filter}\NormalTok{(}\FunctionTok{n}\NormalTok{() }\SpecialCharTok{\textgreater{}} \DecValTok{1}\NormalTok{)}
\end{Highlighting}
\end{Shaded}

As mentioned before, the first step in creating a temporal network is
creating a static base network (base network) which carries all the
information about the network, e.g., the nodes, edges as well as their
attributes. The base network is typically a static weighted network.
Here we define the base network file (the weighted edge file we created
before), we use \texttt{directed=TRUE} to create our network as directed
and we tell the \texttt{network} function that the vertices attributes
are in the net\_nodes file.

\begin{Shaded}
\begin{Highlighting}[]
\NormalTok{NetworkD }\OtherTok{\textless{}{-}} \FunctionTok{network}\NormalTok{(net\_edges\_NLW, }\AttributeTok{directed =} \ConstantTok{TRUE}\NormalTok{, }\AttributeTok{matrix.type =} \StringTok{"edgelist"}\NormalTok{, }
                    \AttributeTok{loops =} \ConstantTok{FALSE}\NormalTok{, }\AttributeTok{multiple =} \ConstantTok{FALSE}\NormalTok{, }\AttributeTok{vertices =}\NormalTok{ net\_nodes)}
\end{Highlighting}
\end{Shaded}

For creating a temporal network, we need more than the source and the
target commonly needed for the static network. In particular, the
following variables are required to be defined.

\begin{itemize}
\item
  \texttt{tail}: the source of the interaction
\item
  \texttt{head}: the target of the interaction
\item
  \texttt{onset}: The starting time of the interaction
\item
  \texttt{terminus}: the end time of the interaction
\item
  \texttt{duration}: the duration of the interaction
\end{itemize}

\begin{figure}

{\centering \includegraphics[width=6.5in,height=2.18056in]{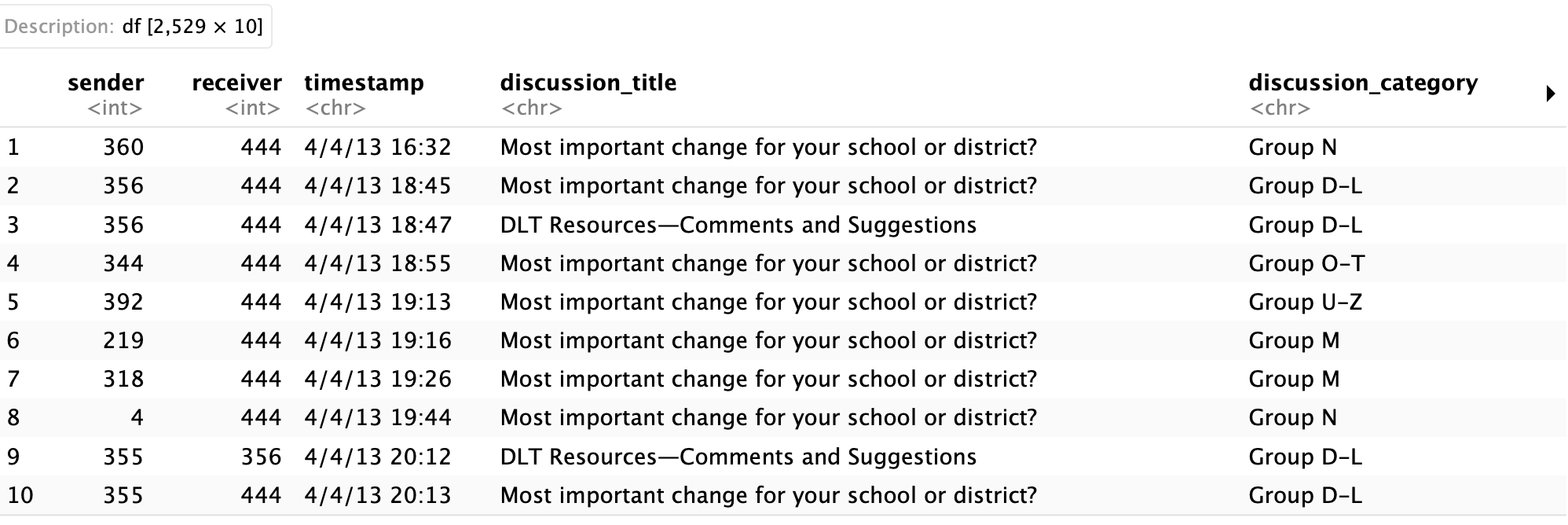}

}

\caption{\label{fig-edges-file}Screenshot of the edges file. The edge
file has source (sender), target (receiver) and a timestamp}

\end{figure}

Our dataset ---which comes from forum MOOC interactions, see
Figure~\ref{fig-edges-file}--- has an obvious starting time (which is
the timestamp of each interaction) but has no clear end time. There is
no straightforward answer to this question. Nonetheless, a possible way
to consider the duration of every post is the duration the post was
active in the discussion or continued to be discussed. That is the time
from the post in a discussion thread to the last post in the same
threads of replies. Such a method ---while far from perfect--- offers a
rough method for estimating the time during which this interaction has
been ``active'' in the discussion {[}15, 22{]}. For an illustration, see
Figure~\ref{fig-edges-file} which shows the durations for the first and
second posts. The next code chunk creates a variable for the starting
time of each interaction, computes the ending time where this post was
part of an active discussion, and then computes the duration.

\begin{figure}

{\centering \includegraphics[width=5.24306in,height=2.8904in]{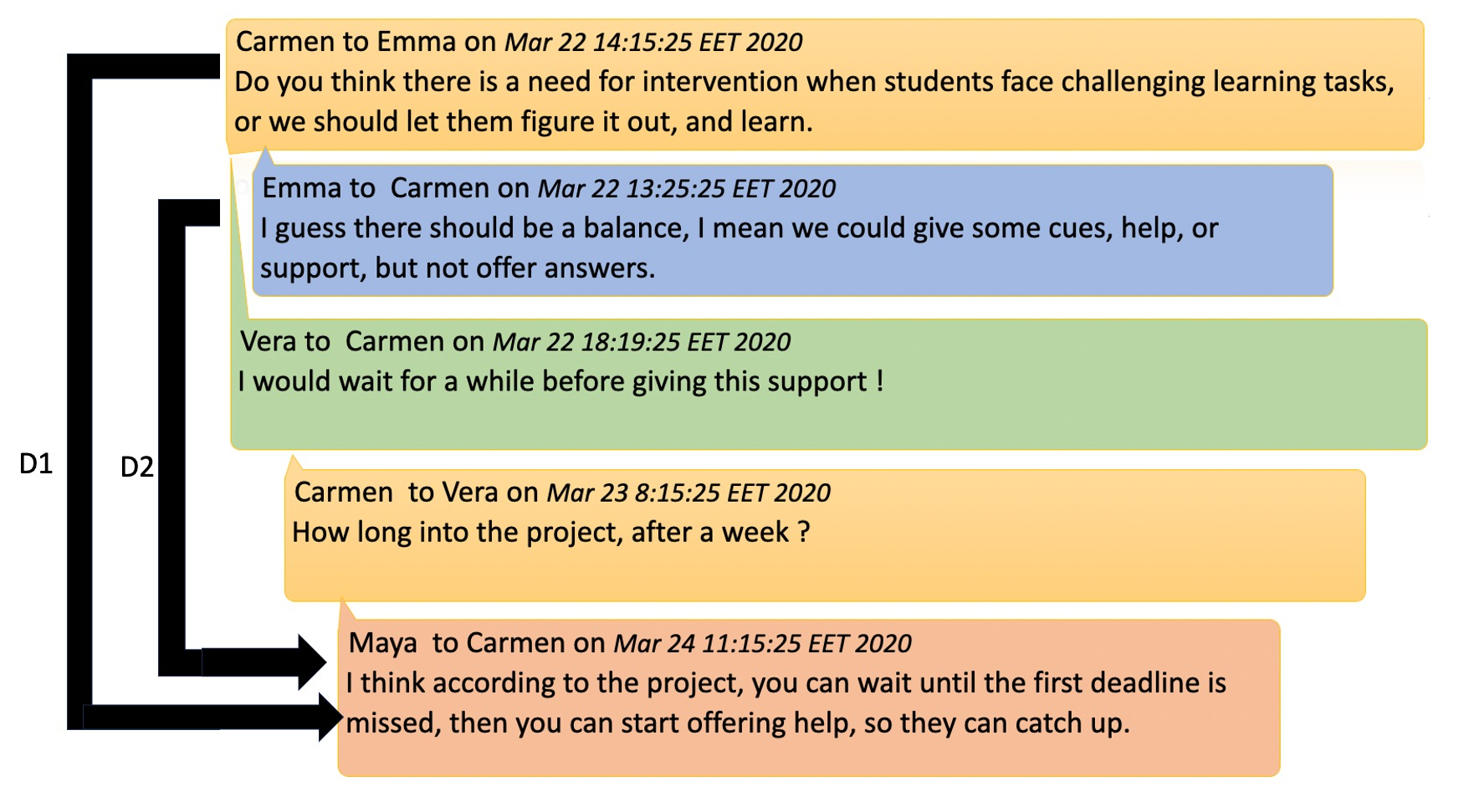}

}

\caption{A sample discussion demonstrating a way to compute the duration
of the post. The first post lasts active in the discussion from starting
time to the last reply it stimulated in the same thread (D1)}

\end{figure}

In the same way, the second duration is computed in the same way (D2).
The next step of the analysis creates a dataframe where all the needed
information for the network is specified and in the next step we simply
use the \texttt{networkDynamic} with two arguments, the base network and
the dataframe with all the temporal network information created in the
previous step. Nonetheless, dealing with around 450 nodes in a network
is hard, it becomes impossible to visualize or get insights from a large
number of crowded nodes. So, for the sake of simplicity of demonstration
in this tutorial, we will create a smaller subset of the network of
people who had a reasonable number of interactions (degree more than 20)
using \texttt{get.inducedSubgraph} argument. The resulting network is
then called \texttt{Active\_Network} which we will analyze.

\begin{Shaded}
\begin{Highlighting}[]
\CommentTok{\# Create the required variables (start, end, and duration) defined by}

\NormalTok{net\_edges\_NL }\OtherTok{\textless{}{-}}\NormalTok{ net\_edges\_NL }\SpecialCharTok{\%\textgreater{}\%} \FunctionTok{group\_by}\NormalTok{(discussion\_title) }\SpecialCharTok{\%\textgreater{}\%}
\FunctionTok{mutate}\NormalTok{(}\AttributeTok{start =} \FunctionTok{min}\NormalTok{(new\_date), }\AttributeTok{end =} \FunctionTok{max}\NormalTok{(new\_date), }\AttributeTok{duration =}\NormalTok{ end}\SpecialCharTok{{-}}\NormalTok{start)}


\CommentTok{\# Creating a dataframe with needed variables}

\NormalTok{edge\_spells }\OtherTok{\textless{}{-}} \FunctionTok{data.frame}\NormalTok{(}\StringTok{"onset"} \OtherTok{=}\NormalTok{ net\_edges\_NL}\SpecialCharTok{\$}\NormalTok{start,}
                         \StringTok{"terminus"} \OtherTok{=}\NormalTok{ net\_edges\_NL}\SpecialCharTok{\$}\NormalTok{end, }
                         \StringTok{"tail"} \OtherTok{=}\NormalTok{ net\_edges\_NL}\SpecialCharTok{\$}\NormalTok{sender, }
                         \StringTok{"head"} \OtherTok{=}\NormalTok{ net\_edges\_NL}\SpecialCharTok{\$}\NormalTok{receiver, }
                         \StringTok{"onset.censored"} \OtherTok{=} \ConstantTok{FALSE}\NormalTok{,}
                         \StringTok{"terminus.censored"} \OtherTok{=} \ConstantTok{FALSE}\NormalTok{, }
                         \StringTok{"duration"} \OtherTok{=}\NormalTok{ net\_edges\_NL}\SpecialCharTok{\$}\NormalTok{duration)}

\CommentTok{\# Creating the dynamic network network}
\NormalTok{Dynamic\_network }\OtherTok{\textless{}{-}} \FunctionTok{networkDynamic}\NormalTok{(NetworkD, }\AttributeTok{edge.spells =}\NormalTok{ edge\_spells)}
\end{Highlighting}
\end{Shaded}

\begin{verbatim}
Edge activity in base.net was ignored
Created net.obs.period to describe network
 Network observation period info:
  Number of observation spells: 1 
  Maximal time range observed: 0 until 72.01 
  Temporal mode: continuous 
  Time unit: unknown 
  Suggested time increment: NA 
\end{verbatim}

\begin{Shaded}
\begin{Highlighting}[]
\NormalTok{Active\_Network }\OtherTok{\textless{}{-}} \FunctionTok{get.inducedSubgraph}\NormalTok{(Dynamic\_network,}
                                \AttributeTok{v =} \FunctionTok{which}\NormalTok{(}\FunctionTok{degree}\NormalTok{(Dynamic\_network) }\SpecialCharTok{\textgreater{}} \DecValTok{20}\NormalTok{))}
\end{Highlighting}
\end{Shaded}

\begin{figure}[H]

{\centering \includegraphics{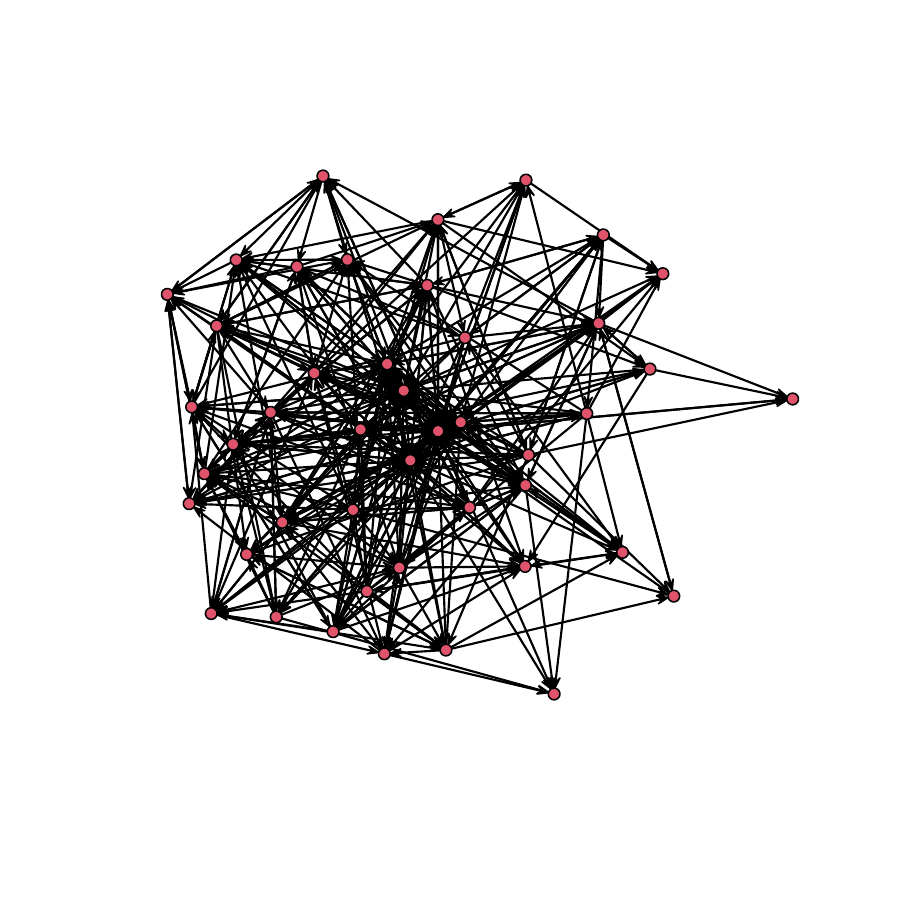}

}

\caption{\label{fig-full-network}A plot of the full network as plotted
by the plot function.}

\end{figure}

We can then confirm that the network has been created correctly using
the \texttt{print} function. As the output shows, we have 521 distinct
time changes, 72 days, 445 vertices and 1936 edges. We can also use the
function \texttt{plot} to see how the network looks. The argument
\texttt{pad} helps us remove the additional whitespace around the
network. Plotting a temporal network helps summarize all the
interactions in the network. As we can see in
Figure~\ref{fig-full-network}, the network is dense with several edges
between interacting students.

\begin{Shaded}
\begin{Highlighting}[]
\FunctionTok{print}\NormalTok{(Dynamic\_network)}
\end{Highlighting}
\end{Shaded}

\begin{verbatim}
NetworkDynamic properties:
  distinct change times: 495 
  maximal time range: 0 until  72.01 

Includes optional net.obs.period attribute:
 Network observation period info:
  Number of observation spells: 1 
  Maximal time range observed: 0 until 72.01 
  Temporal mode: continuous 
  Time unit: unknown 
  Suggested time increment: NA 

 Network attributes:
  vertices = 445 
  directed = TRUE 
  hyper = FALSE 
  loops = FALSE 
  multiple = FALSE 
  bipartite = FALSE 
  net.obs.period: (not shown)
  total edges= 1936 
    missing edges= 0 
    non-missing edges= 1936 

 Vertex attribute names: 
    connect country experience experience2 expert expert_level Facilitator gender grades group location region role1 vertex.names 

 Edge attribute names not shown 
\end{verbatim}

\begin{Shaded}
\begin{Highlighting}[]
\FunctionTok{plot.network}\NormalTok{(Active\_Network, }\AttributeTok{pad =} \SpecialCharTok{{-}}\FloatTok{0.5}\NormalTok{)}
\end{Highlighting}
\end{Shaded}

\hypertarget{visualization-of-temporal-networks}{%
\subsection{Visualization of temporal
networks}\label{visualization-of-temporal-networks}}

\begin{figure}

{\centering \includegraphics{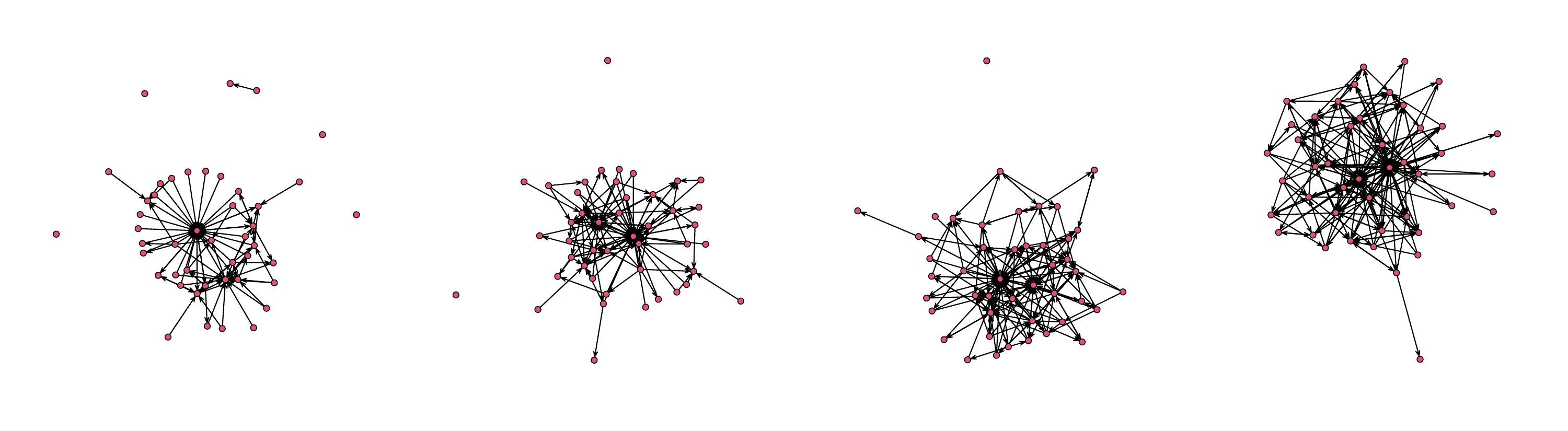}

}

\caption{\label{fig-plot-weeks}A plot of the network at weeks 1, 2, 3,
and 4}

\end{figure}

\begin{figure}[H]

{\centering \includegraphics{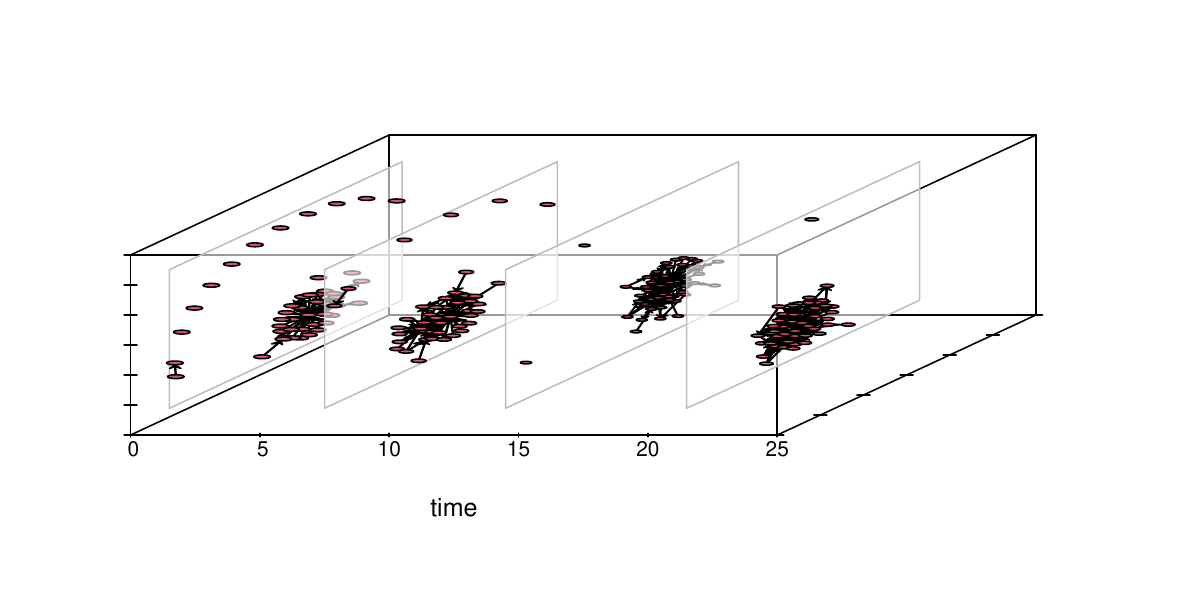}

}

\caption{\label{fig-timeprism}A three-dimensional visualization of the
network as a sequence of snapshots in space and time prism}

\end{figure}

To take advantage of the temporal network, we can use a function to
extract the network at certain times to explore the activity. In the
next example in Figure~\ref{fig-plot-weeks}, we chose the first four
weeks one by one and plotted them alongside each other. The function
\texttt{filmstrip} can create a similar output with a snapshot of the
network at several time intervals. A similar result, yet with
three-dimensional placing, can also be obtained with the
\texttt{timePrism} function, as shown in Figure~\ref{fig-timeprism}. You
may need to consult the package manual to get more information about the
arguments and options for the plots.

\begin{Shaded}
\begin{Highlighting}[]
\FunctionTok{plot.network}\NormalTok{(}\FunctionTok{network.extract}\NormalTok{(Active\_Network, }\AttributeTok{onset =} \DecValTok{1}\NormalTok{, }\AttributeTok{terminus =} \DecValTok{7}\NormalTok{))}
\FunctionTok{plot.network}\NormalTok{(}\FunctionTok{network.extract}\NormalTok{(Active\_Network, }\AttributeTok{onset =} \DecValTok{8}\NormalTok{, }\AttributeTok{terminus =} \DecValTok{14}\NormalTok{))}
\FunctionTok{plot.network}\NormalTok{(}\FunctionTok{network.extract}\NormalTok{(Active\_Network, }\AttributeTok{onset =} \DecValTok{15}\NormalTok{, }\AttributeTok{terminus =} \DecValTok{21}\NormalTok{))}
\FunctionTok{plot.network}\NormalTok{(}\FunctionTok{network.extract}\NormalTok{(Active\_Network, }\AttributeTok{onset =} \DecValTok{22}\NormalTok{, }\AttributeTok{terminus =} \DecValTok{28}\NormalTok{))}
\end{Highlighting}
\end{Shaded}

\begin{Shaded}
\begin{Highlighting}[]
\FunctionTok{compute.animation}\NormalTok{(Active\_Network)}
\end{Highlighting}
\end{Shaded}

\begin{verbatim}
slice parameters:
  start:0
  end:72.01
  interval:1
  aggregate.dur:1
  rule:latest
\end{verbatim}

\begin{Shaded}
\begin{Highlighting}[]
\FunctionTok{timePrism}\NormalTok{(Active\_Network, }\AttributeTok{at =} \FunctionTok{c}\NormalTok{(}\DecValTok{1}\NormalTok{, }\DecValTok{7}\NormalTok{, }\DecValTok{14}\NormalTok{, }\DecValTok{21}\NormalTok{),}
  \AttributeTok{spline.lwd =} \DecValTok{1}\NormalTok{,}
  \AttributeTok{box =} \ConstantTok{TRUE}\NormalTok{,}
  \AttributeTok{angle =} \DecValTok{60}\NormalTok{,}
  \AttributeTok{axis =} \ConstantTok{TRUE}\NormalTok{,}
  \AttributeTok{planes =} \ConstantTok{TRUE}\NormalTok{,}
  \AttributeTok{plane.col =} \StringTok{"\#FFFFFF99"}\NormalTok{,}
  \AttributeTok{scale.y =} \DecValTok{1}\NormalTok{,}
  \AttributeTok{orientation =} \FunctionTok{c}\NormalTok{(}\StringTok{"z"}\NormalTok{, }\StringTok{"x"}\NormalTok{, }\StringTok{"y"}\NormalTok{))}
\end{Highlighting}
\end{Shaded}

\begin{figure}

{\centering \includegraphics[width=4.64063in,height=2.87808in]{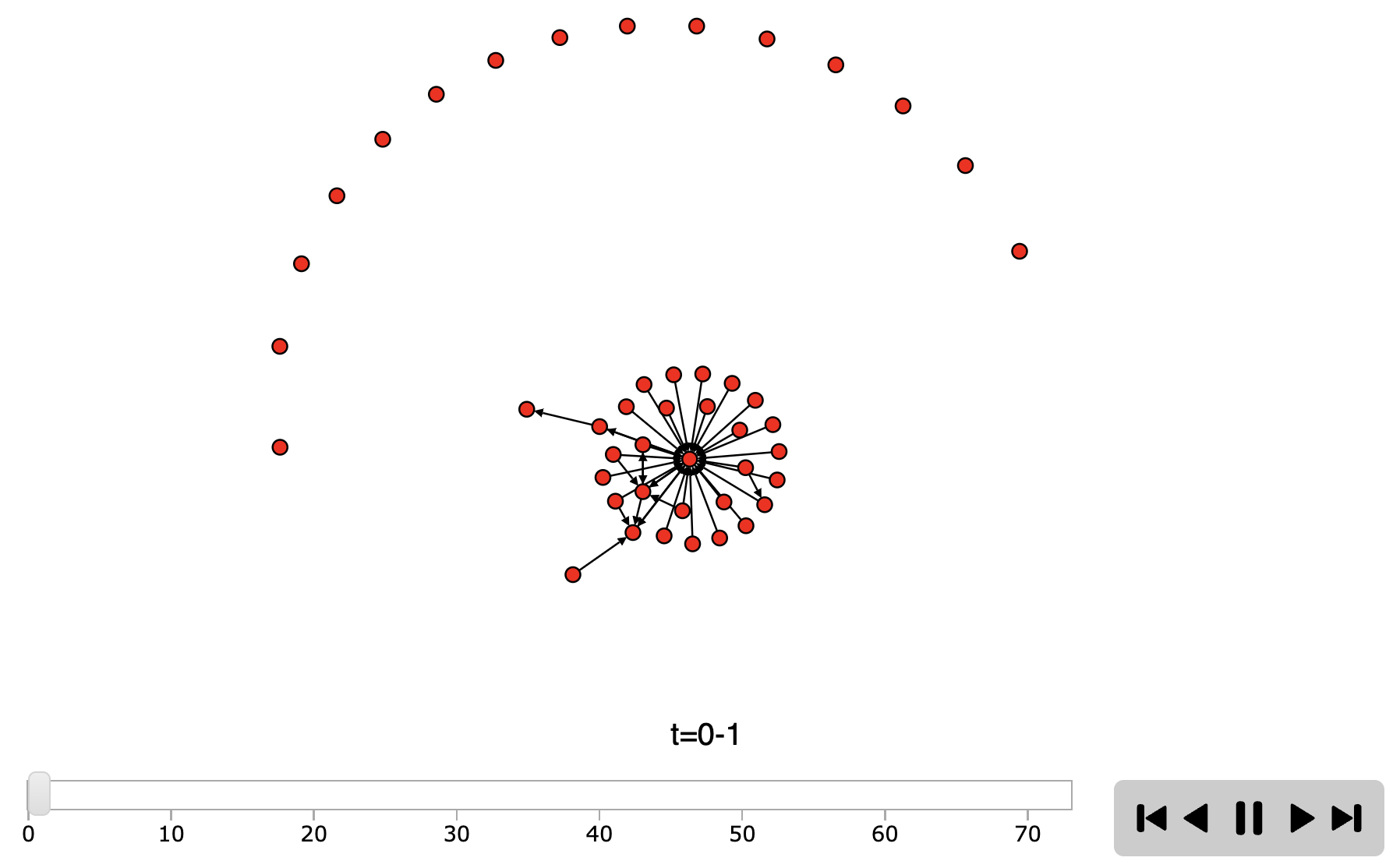}

}

\caption{\label{fig-movie}A screenshot of the animated movie of the
temporal network}

\end{figure}

However, a better way is to take advantage of the capabilities of the
package \texttt{ndtv} and the network temporal information by rendering
a full animated movie of the network as in Figure~\ref{fig-movie} and
explore each and every event as it happens.

\begin{Shaded}
\begin{Highlighting}[]
\FunctionTok{render.d3movie}\NormalTok{(Active\_Network)}
\end{Highlighting}
\end{Shaded}

\begin{figure}

\begin{minipage}[t]{0.50\linewidth}

{\centering 

\raisebox{-\height}{

\includegraphics{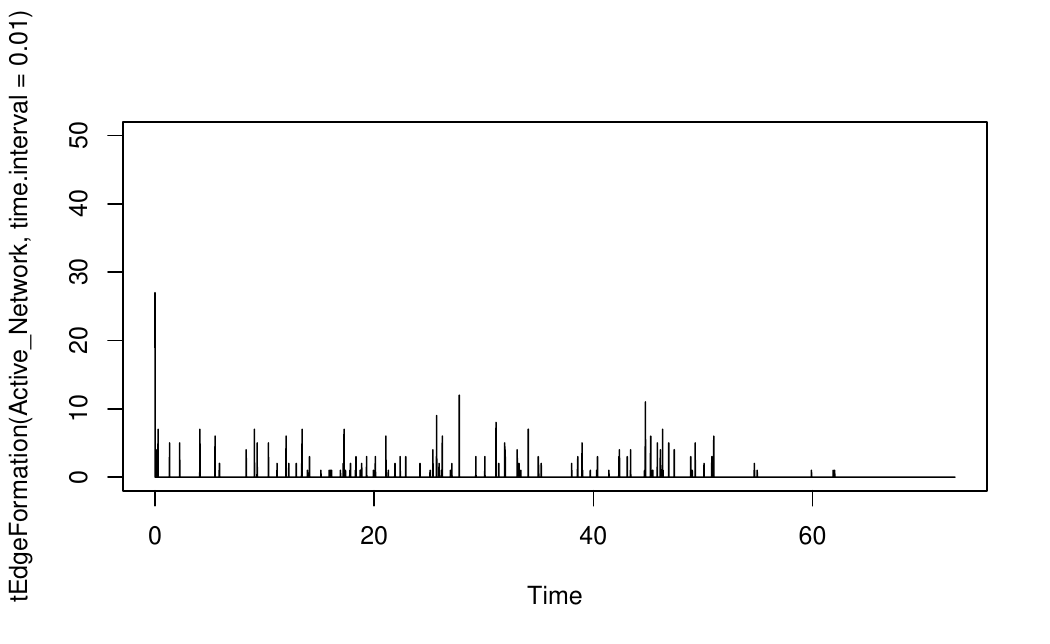}

}

}

\subcaption{\label{fig-edges-1}Edge formation}
\end{minipage}%
\begin{minipage}[t]{0.50\linewidth}

{\centering 

\raisebox{-\height}{

\includegraphics{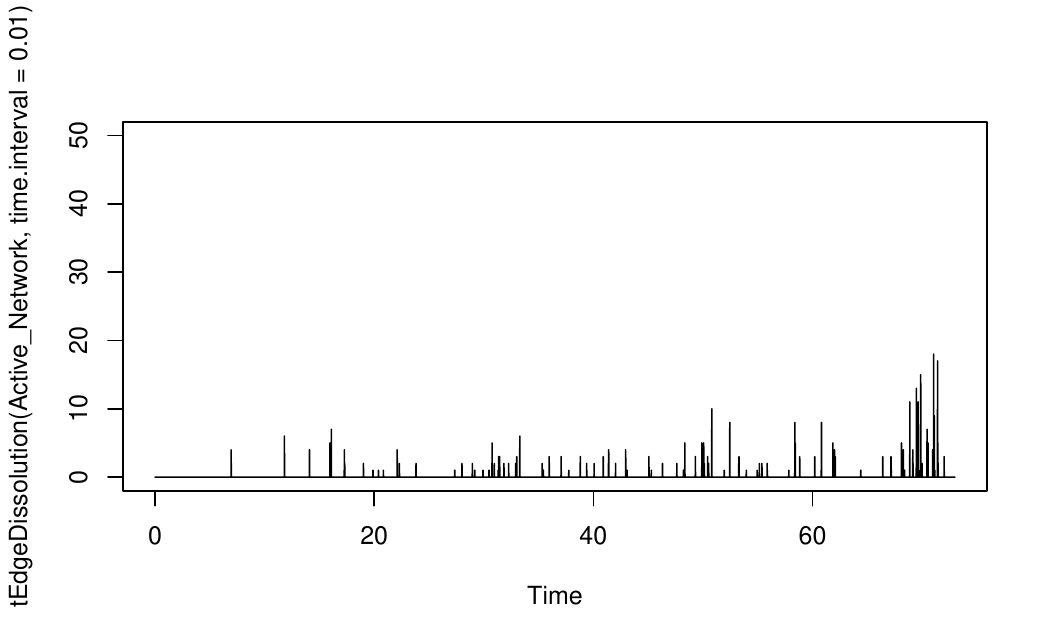}

}

}

\subcaption{\label{fig-edges-2}Edge dissolution}
\end{minipage}%

\caption{\label{fig-edges}Edge formation and dissolution. We see that
edge formation occurs towards the start and increases till almost t=50
and edge dissolution starts scanty at the beginning and increases with
time, peaking at t=70}

\end{figure}

As we mentioned in the introduction section, in temporal networks, edges
or relationships form ``get activated'' and dissolve ``get
deactivated''. We can plot such the dynamic edge formation and
dissolution process using the functions \texttt{tEdgeFormation} which as
the name implies plots the edges forming at the given time point. The
function \texttt{tEdgeDissolution} returns the edges terminating and can
be plotted in the same way as seen in Figure~\ref{fig-edges}. Obviously,
at the beginning of the MOOC, we see new relationships form and, at the
end, most relationships dissolve.

\begin{Shaded}
\begin{Highlighting}[]
\FunctionTok{plot}\NormalTok{(}\FunctionTok{tEdgeFormation}\NormalTok{(Active\_Network, }\AttributeTok{time.interval =} \FloatTok{0.01}\NormalTok{), }\AttributeTok{ylim =} \FunctionTok{c}\NormalTok{(}\DecValTok{0}\NormalTok{,}\DecValTok{50}\NormalTok{))}
\FunctionTok{plot}\NormalTok{(}\FunctionTok{tEdgeDissolution}\NormalTok{(Active\_Network, }\AttributeTok{time.interval =} \FloatTok{0.01}\NormalTok{), }\AttributeTok{ylim =} \FunctionTok{c}\NormalTok{(}\DecValTok{0}\NormalTok{, }\DecValTok{50}\NormalTok{))}
\end{Highlighting}
\end{Shaded}

\begin{figure}[H]

{\centering \includegraphics{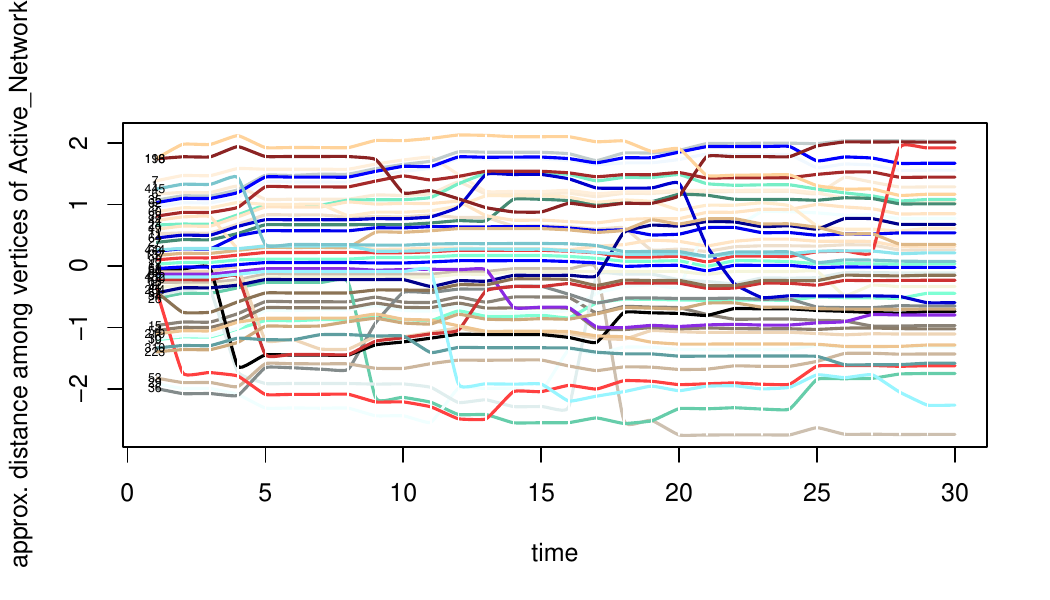}

}

\caption{\label{fig-prox-timeline}A proximity timeline shows connected
interacting nodes closer to each other.}

\end{figure}

Another way to visualize a temporal network is to use the proximity
timeline, the \texttt{proximity.timeline} function tries to draw the
temporal network in two dimensions, that is, it draws nodes at each time
point taking into account how closely connected they are and renders
them accordingly, nodes that are interacting are rendered close to each
other, and nodes that not interacting are rendered apart. Technically as
described in the function manual: ``The passed network dynamic object is
sliced up into a series of networks. It loops over the networks,
converting each to a distance matrix based on geodesic path distance
with \texttt{layout.distance}. The distances are fed into an MDS
algorithm (specified by mode) that lays them out in one dimension:
essentially trying to position them along a vertical line. The sequence
of 1D layouts are arranged along a timeline, and a spline is drawn for
each vertex connecting its positions at each time point. The idea is
that closely-linked clusters form bands of lines that move together
through the plot'' {[}20{]}. The result is a timeline of temporal
proximity. The next code draws the proximity timeline but also adds some
colors, and a start and end for the plot, see the result in
Figure~\ref{fig-prox-timeline}.

\begin{Shaded}
\begin{Highlighting}[]
\FunctionTok{proximity.timeline}\NormalTok{(Active\_Network, }\AttributeTok{default.dist =} \DecValTok{1}\NormalTok{, }\AttributeTok{mode =} \StringTok{"sammon"}\NormalTok{,}
                   \AttributeTok{labels.at =} \DecValTok{1}\NormalTok{, }\AttributeTok{vertex.col =}\NormalTok{ grDevices}\SpecialCharTok{::}\FunctionTok{colors}\NormalTok{(),}
                   \AttributeTok{start =} \DecValTok{1}\NormalTok{, }\AttributeTok{end =} \DecValTok{30}\NormalTok{, }\AttributeTok{label.cex =} \FloatTok{0.5}\NormalTok{)}
\end{Highlighting}
\end{Shaded}

\hypertarget{statistical-analysis-of-temporal-networks}{%
\subsection{Statistical analysis of temporal
networks}\label{statistical-analysis-of-temporal-networks}}

\hypertarget{graph-level-measures}{%
\subsubsection{Graph level measures}\label{graph-level-measures}}

Graph properties in temporal networks are dynamic and vary by time. When
graph measures are computed we get a time series of the computed
measures. Such fine-grained measures allow us to understand how the
networks and their structure evolve or unfold in time and thus, such
information can help us understand collaboration or interaction dynamics
as they occur {[}9, 15{]}.

The function \texttt{tSnaStats} from the package \texttt{tsna} has a
large number of measures that can be computed by specifying the argument
\texttt{snafun}. In the next example, we compute the graph level density
with the argument \texttt{snafun=gden.} The function allows the choice
of a range of time, for instance, from the end of second week to the end
of second month by supplying the arguments start=14 to end=60. We can
also use the \texttt{time.interval} to specify the granularity of the
calculation. The argument \texttt{aggregate.dur} specifies the period of
the aggregation, for instance, \texttt{aggregate.dur=7} will compute the
density for every seven days. We can also plot the density time series
by simply using the function \texttt{plot}.

\begin{Shaded}
\begin{Highlighting}[]
\NormalTok{Density }\OtherTok{\textless{}{-}} \FunctionTok{tSnaStats}\NormalTok{(}
  \AttributeTok{nd =}\NormalTok{ Active\_Network,}
  \AttributeTok{snafun =} \StringTok{"gden"}\NormalTok{,}
  \AttributeTok{start =} \DecValTok{14}\NormalTok{,}
  \AttributeTok{end =} \DecValTok{60}\NormalTok{,}
  \AttributeTok{time.interval =} \DecValTok{1}\NormalTok{,}
  \AttributeTok{aggregate.dur =} \DecValTok{7}\NormalTok{)}
\FunctionTok{plot}\NormalTok{(Density)}
\end{Highlighting}
\end{Shaded}

\begin{figure}[H]

{\centering \includegraphics{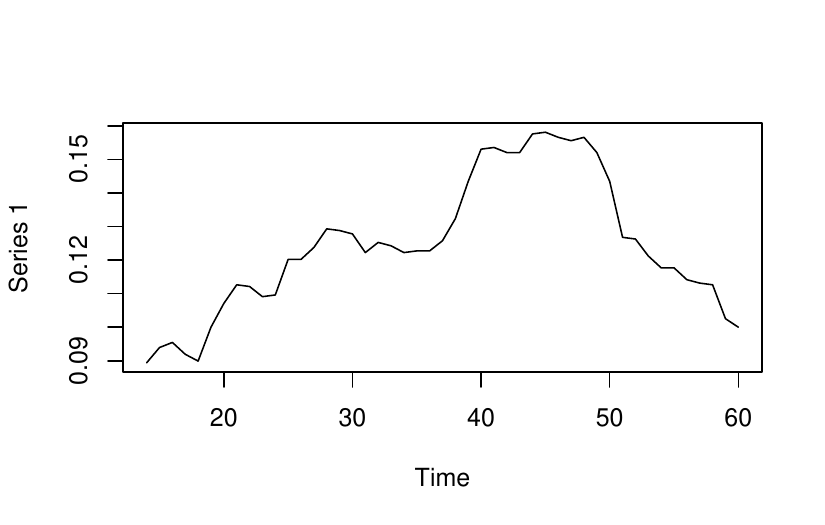}

}

\caption{\label{fig-temporal-density}A plot of temporal density from
t=14 to t=60}

\end{figure}

You can see, in the resulting graph in
Figure~\ref{fig-temporal-density}, that the density increases until day
50 and then starts to drop. Of note, another type of density can be
computed, known as temporal density, which computes the observed total
duration of all edges and divides it by the maximum duration possible.
Temporal density can be computed using the command
\texttt{tEdgeDensity}.

\begin{Shaded}
\begin{Highlighting}[]
\FunctionTok{tEdgeDensity}\NormalTok{(Active\_Network) }
\end{Highlighting}
\end{Shaded}

\begin{verbatim}
[1] 0.3901841
\end{verbatim}

\begin{Shaded}
\begin{Highlighting}[]
\FunctionTok{gden}\NormalTok{(Active\_Network) }
\end{Highlighting}
\end{Shaded}

\begin{verbatim}
[1] 0.2186869
\end{verbatim}

\begin{figure}

\begin{minipage}[t]{0.33\linewidth}

{\centering 

\raisebox{-\height}{

\includegraphics{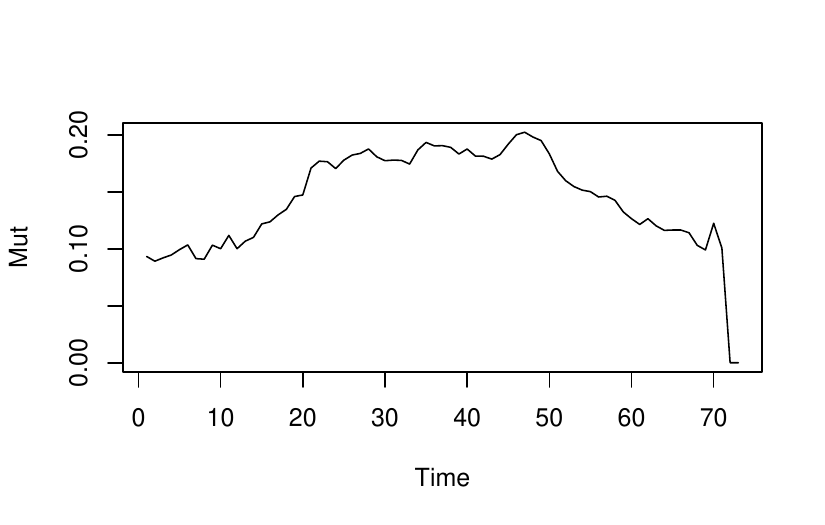}

}

}

\subcaption{\label{fig-several-1}}
\end{minipage}%
\begin{minipage}[t]{0.33\linewidth}

{\centering 

\raisebox{-\height}{

\includegraphics{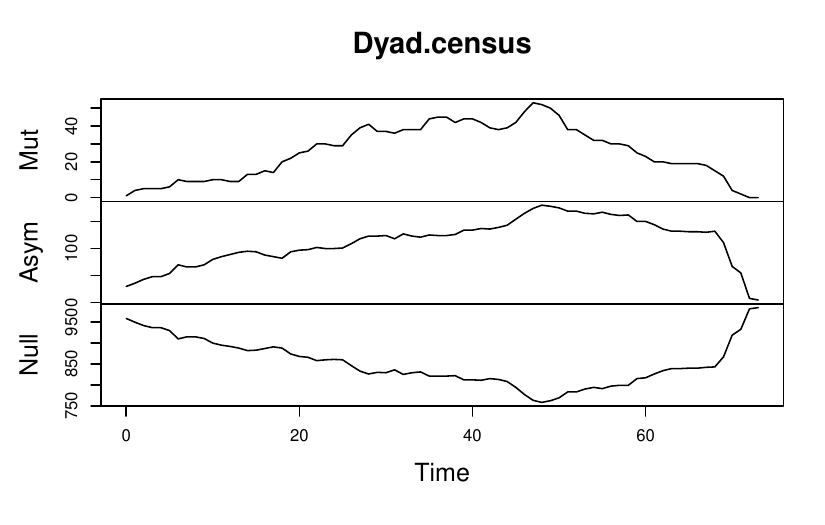}

}

}

\subcaption{\label{fig-several-2}}
\end{minipage}%
\begin{minipage}[t]{0.33\linewidth}

{\centering 

\raisebox{-\height}{

\includegraphics{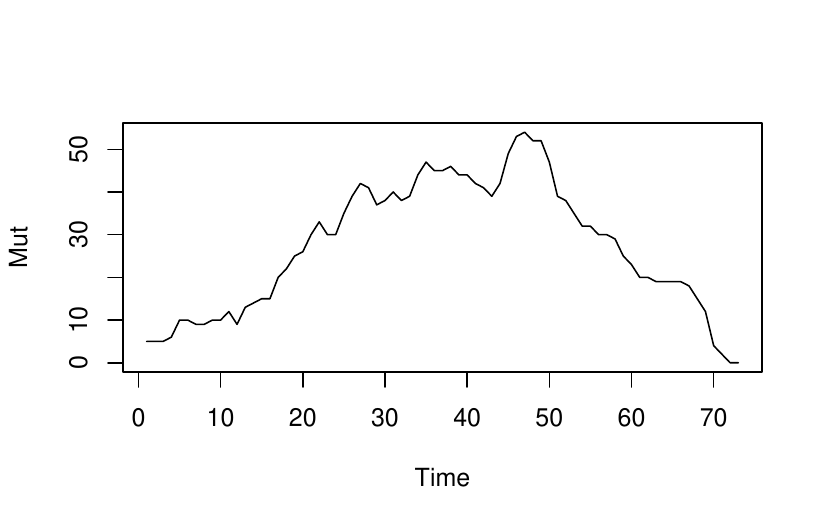}

}

}

\subcaption{\label{fig-several-3}}
\end{minipage}%

\caption{\label{fig-several} (A) Reciprocity over time. (B) Dyad census
over time. (C) Mutuality over time.}

\end{figure}

Similar to density, we can compute reciprocity, i.e., the ratio of
reciprocated edges to asymmetric edges. Note, that we calculate
reciprocity here from day 1 to day 73 on a daily basis (this is just for
demonstration of different periods). Since the function default is to
calculate the reciprocated dyads, we specify the argument \emph{measure
= ``edgewise''} to calculate the proportion of reciprocated edges. As
the graph in Figure~\ref{fig-several} (a) shows, reciprocity increases
steadily for the first 50 days pointing to a build up of trust between
collaborators. The \texttt{dyad.census} function offers a more granular
view of the dyads and their reciprocity as shown in
Figure~\ref{fig-several} (b). Similarly, \texttt{mutuality} is a very
close function and returns the number of complete dyads (reciprocated
dyads), plotted in Figure~\ref{fig-several} (c). All of the
aforementioned functions deal with reciprocity in one way or another,
for differences and usages, readers are encouraged to read the
functions' help to explore the differences, arguments as well as the
equation for each function.

\begin{Shaded}
\begin{Highlighting}[]
\NormalTok{Reciprocity }\OtherTok{\textless{}{-}} \FunctionTok{tSnaStats}\NormalTok{(}
    \AttributeTok{nd=}\NormalTok{Dynamic\_network,}
    \AttributeTok{snafun =} \StringTok{"grecip"}\NormalTok{ ,}
    \AttributeTok{start =} \DecValTok{1}\NormalTok{,}
    \AttributeTok{end =} \DecValTok{73}\NormalTok{,}
    \AttributeTok{measure =}\StringTok{"edgewise"}\NormalTok{,}
    \AttributeTok{time.interval =} \DecValTok{1}\NormalTok{,}
    \AttributeTok{aggregate.dur =} \DecValTok{1}\NormalTok{)}
\FunctionTok{plot}\NormalTok{(Reciprocity)}

\NormalTok{Dyad.census }\OtherTok{\textless{}{-}} \FunctionTok{tSnaStats}\NormalTok{(Active\_Network, }
                         \AttributeTok{snafun =}\StringTok{"dyad.census"}\NormalTok{)}
\FunctionTok{plot}\NormalTok{(Dyad.census)}

\NormalTok{dynamicmutuality }\OtherTok{\textless{}{-}} \FunctionTok{tSnaStats}\NormalTok{(}
\NormalTok{  Active\_Network,}
  \AttributeTok{snafun =} \StringTok{"mutuality"}\NormalTok{,}
  \AttributeTok{start =} \DecValTok{1}\NormalTok{,}
  \AttributeTok{end =} \DecValTok{73}\NormalTok{,}
  \AttributeTok{time.interval =} \DecValTok{1}\NormalTok{,}
  \AttributeTok{aggregate.dur =} \DecValTok{1}
\NormalTok{  )}
\FunctionTok{plot}\NormalTok{(dynamicmutuality)}
\end{Highlighting}
\end{Shaded}

\begin{figure}[H]

{\centering \includegraphics{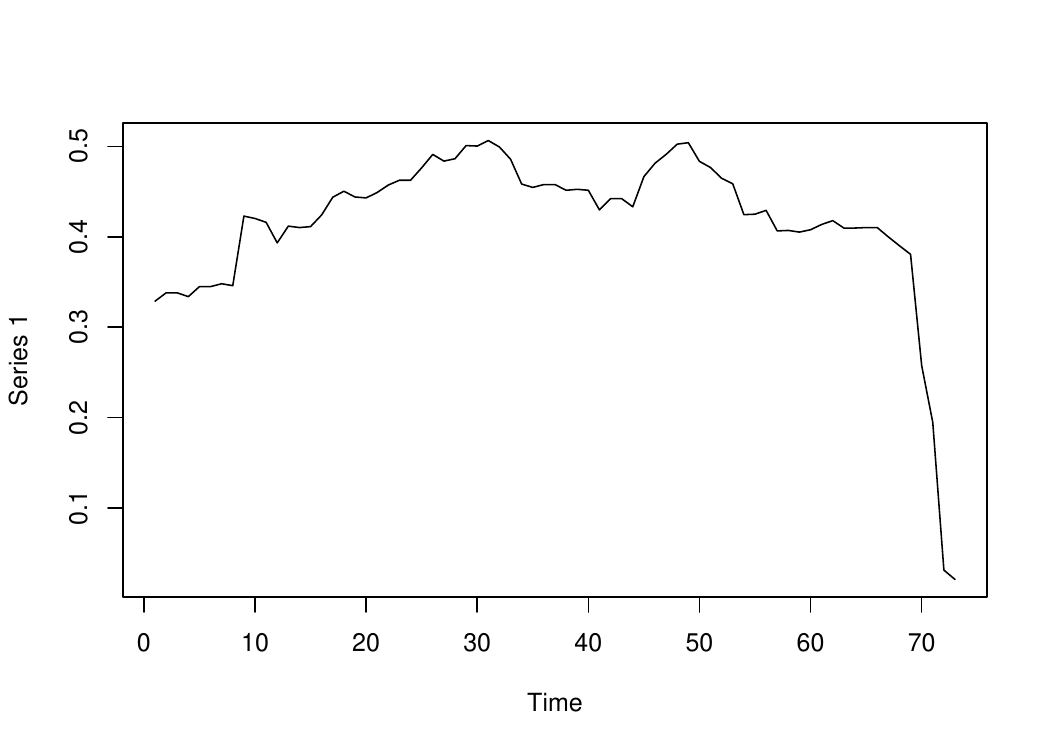}

}

\caption{\label{fig-deg-centr}A plot of degree centralization over time}

\end{figure}

Centralization measures the dominance of members within the network and
can be traced temporally using the function snafun = ``centralization''.
Please note that we can choose the period, the interval and the
aggregation periods using function arguments as mentioned before. The
next example computes the degree centralization as demonstrated in
Figure~\ref{fig-deg-centr}. Also note that we can also compute other
centralization measures such as centralization indegree, centralization
outdegree, centralization betweenness, centralization closeness and
eigenvector.

\begin{Shaded}
\begin{Highlighting}[]
\NormalTok{Degree\_centralization }\OtherTok{\textless{}{-}} \FunctionTok{tSnaStats}\NormalTok{(}
\NormalTok{  Active\_Network,}
  \AttributeTok{snafun =} \StringTok{"centralization"}\NormalTok{,}
  \AttributeTok{start =} \DecValTok{1}\NormalTok{,}
  \AttributeTok{end =} \DecValTok{73}\NormalTok{,}
  \AttributeTok{time.interval =} \DecValTok{1}\NormalTok{,}
  \AttributeTok{aggregate.dur =} \DecValTok{1}\NormalTok{,}
  \AttributeTok{FUN =} \StringTok{"degree"}\NormalTok{)}

\FunctionTok{plot}\NormalTok{(Degree\_centralization)}
\end{Highlighting}
\end{Shaded}

Several other graph-level measure can be computed in the same way using
the following arguments passed to the \texttt{snafun}: -
\texttt{components}: count of Components within the graph over time -
\texttt{triad.census}: the triad census and types of triads over time -
\texttt{connectedness}: the connectedness score of the network
efficiency network efficiency over time - \texttt{gtrans}: network
transitivity over time - \texttt{hierarchy}: network Hierarchy over time
- \texttt{lubness}: network LUBness over time - \texttt{efficiency}:
network efficiency over time - \texttt{hierarchy}: network hierarchy
over time

\begin{figure}[H]

{\centering \includegraphics{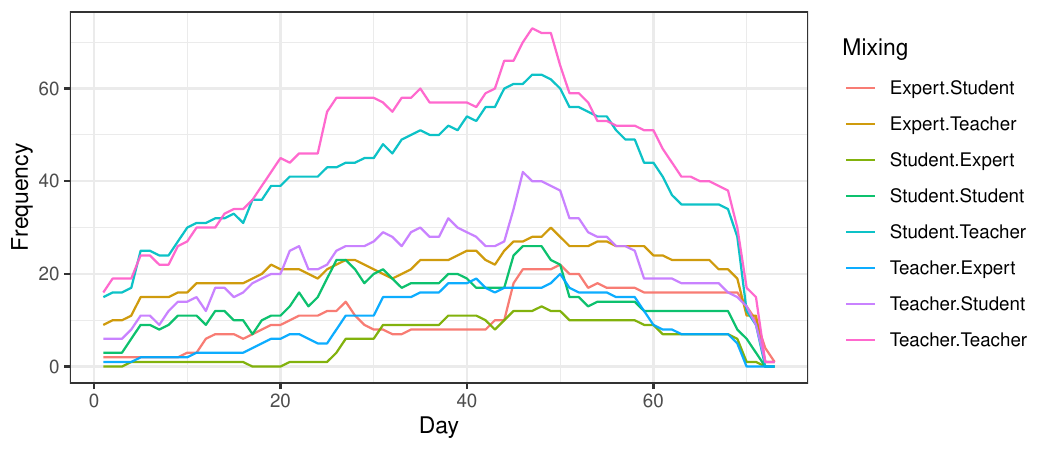}

}

\caption{\label{fig-mix-exp}Mixing patterns between expertise levels and
the teachers.}

\end{figure}

Another useful package that offers a wealth of graph level measures is
\texttt{tErgmStats}, you may need to consult the help files which are
available by using the command \texttt{?tErgmStats}. One of the
important functions that we can try here is the \texttt{nodemix}. We can
use nodemix to examine who and when different actors interact with each
other, see here for an example where the authors examined how low and
high achievers mix with each other and with the teachers {[}15{]}. The
next code demonstrates how to compute the mixing patterns between
expertise levels. We then convert the time series as a dataframe, clean
the names and then plot the results as demonstrated in
Figure~\ref{fig-mix-exp}.

\begin{Shaded}
\begin{Highlighting}[]
\NormalTok{Mix\_experience}\OtherTok{\textless{}{-}} \FunctionTok{tErgmStats}\NormalTok{(Active\_Network,}
  \StringTok{"nodemix(\textquotesingle{}expert\_level\textquotesingle{})"}\NormalTok{,}
  \AttributeTok{start =} \DecValTok{1}\NormalTok{,}
  \AttributeTok{end =} \DecValTok{73}\NormalTok{,}
  \AttributeTok{time.interval =} \DecValTok{1}\NormalTok{,}
  \AttributeTok{aggregate.dur =} \DecValTok{1}\NormalTok{)}

\NormalTok{Mixing }\OtherTok{\textless{}{-}} \FunctionTok{as.data.frame}\NormalTok{(Mix\_experience)}

\FunctionTok{colnames}\NormalTok{(Mixing) }\OtherTok{\textless{}{-}} \FunctionTok{gsub}\NormalTok{(}\StringTok{"mix.expert\_level."}\NormalTok{, }\StringTok{""}\NormalTok{, }\FunctionTok{colnames}\NormalTok{(Mixing))}

\NormalTok{Mixing}\SpecialCharTok{\$}\NormalTok{Day }\OtherTok{\textless{}{-}} \DecValTok{1}\SpecialCharTok{:}\DecValTok{73}

\NormalTok{Mixing\_long}\OtherTok{=} \FunctionTok{pivot\_longer}\NormalTok{(Mixing, }\FunctionTok{contains}\NormalTok{(}\StringTok{"."}\NormalTok{))}

\FunctionTok{names}\NormalTok{(Mixing\_long)}\OtherTok{=} \FunctionTok{c}\NormalTok{(}\StringTok{"Day"}\NormalTok{, }\StringTok{"Mixing"}\NormalTok{, }\StringTok{"Frequency"}\NormalTok{)}

\FunctionTok{ggplot}\NormalTok{(Mixing\_long,}\FunctionTok{aes}\NormalTok{(Day, Frequency, }\AttributeTok{group=}\NormalTok{Mixing, }\AttributeTok{color=}\NormalTok{Mixing)) }\SpecialCharTok{+} 
  \FunctionTok{geom\_line}\NormalTok{ (}\AttributeTok{alpha=}\NormalTok{.}\DecValTok{95}\NormalTok{) }\SpecialCharTok{+} \FunctionTok{theme\_bw}\NormalTok{()}
\end{Highlighting}
\end{Shaded}

\hypertarget{temporal-centrality-measures}{%
\subsubsection{Temporal centrality
measures}\label{temporal-centrality-measures}}

Centrality measures can be used to identify important actors, as a proxy
indicator for academic achievement or to identify students'
collaborative roles {[}23--26{]}. In temporal networks, centrality
measures are fine-grained estimates of students'' real-time centralities
or importance, i.e., shows who were central and when considering the
temporality of their interaction. In doing so, we can see exactly when
and for how long, at what pace, and with which rhythm a behavior happens
(compare this to traditional social network analysis where centralities
are computed as a single number). There are several possible uses of
temporal centrality measures. For instance {[}15{]} have used them to
create predictive models of students' performance. In another study by
the same authors, they demonstrated that temporal centrality measures
were more predictive of performance compared to traditional static
centrality measures {[}9{]}. Since the temporal centralities are
computed as time series, their temporal characteristics can also be used
to compute other time series properties e.g., stability, variabilities,
and pace, you can see for example {[}15{]}. There is a growing list of
temporal centrality measures e.g., {[}13{]}. We will study here the most
commonly used ones according to the latest review {[}25{]}, but readers
are encouraged to explore the \texttt{tsna} manual for more centrality
measures.

Temporal degree centrality measures can be computed in the same way as
we computed the graph level properties shown before. The next code
defines the function \texttt{snafun\ =\ "degree"}, the start, the end
date, and aggregation (which you can modify). An important argument here
is the \texttt{cmode} argument which defines the type of centrality:
``freeman'', ``indegree'' or ``outdegree'' for the calculation of total
in or out degree centralities. The result is a time series with the 73
values for each day from start to end, each day having a unique value
for the degree centrality for each node. The rest of the code is
intended to organize the results. We convert the time series to a
dataframe, create a variable for the day number to make it easier to
identify the day and create a variable to define the type of centrality,
and then combine all centrality measures into a single data frame as
below.

\begin{Shaded}
\begin{Highlighting}[]
\NormalTok{Degree\_Centrality }\OtherTok{\textless{}{-}} \FunctionTok{tSnaStats}\NormalTok{(}
\NormalTok{  Active\_Network,}
  \AttributeTok{snafun =} \StringTok{"degree"}\NormalTok{,}
  \AttributeTok{start =} \DecValTok{1}\NormalTok{,}
  \AttributeTok{end =} \DecValTok{73}\NormalTok{,}
  \AttributeTok{time.interval =} \DecValTok{1}\NormalTok{,}
  \AttributeTok{aggregate.dur =} \DecValTok{1}\NormalTok{,}
  \AttributeTok{cmode =}\StringTok{"freeman"}\NormalTok{)}

\NormalTok{inDegree\_Centrality }\OtherTok{\textless{}{-}} \FunctionTok{tSnaStats}\NormalTok{(}
\NormalTok{  Active\_Network,}
  \AttributeTok{snafun =} \StringTok{"degree"}\NormalTok{,}
  \AttributeTok{start =} \DecValTok{1}\NormalTok{,}
  \AttributeTok{end =} \DecValTok{73}\NormalTok{,}
  \AttributeTok{time.interval =} \DecValTok{1}\NormalTok{,}
  \AttributeTok{aggregate.dur =} \DecValTok{1}\NormalTok{,}
  \AttributeTok{cmode =}\StringTok{"indegree"}\NormalTok{)}

\NormalTok{OutDegree\_Centrality }\OtherTok{\textless{}{-}} \FunctionTok{tSnaStats}\NormalTok{(}
\NormalTok{  Active\_Network,}
  \AttributeTok{snafun =} \StringTok{"degree"}\NormalTok{,}
  \AttributeTok{start =} \DecValTok{1}\NormalTok{,}
  \AttributeTok{end =} \DecValTok{73}\NormalTok{,}
  \AttributeTok{time.interval =} \DecValTok{1}\NormalTok{,}
  \AttributeTok{aggregate.dur =} \DecValTok{1}\NormalTok{,}
  \AttributeTok{cmode =}\StringTok{"outdegree"}\NormalTok{)}

\NormalTok{Degree\_Centrality }\SpecialCharTok{\%\textgreater{}\%} \FunctionTok{as.data.frame}\NormalTok{() }\SpecialCharTok{\%\textgreater{}\%} \FunctionTok{mutate}\NormalTok{(}\AttributeTok{Day =}
\DecValTok{1}\SpecialCharTok{:}\DecValTok{73}\NormalTok{,}\AttributeTok{centrality=} \StringTok{"Degree\_Centrality"}\NormalTok{, }\AttributeTok{.before=}\NormalTok{1L)}\OtherTok{{-}\textgreater{}}
\NormalTok{Degree\_Centrality\_DF}

\NormalTok{inDegree\_Centrality }\SpecialCharTok{\%\textgreater{}\%} \FunctionTok{as.data.frame}\NormalTok{() }\SpecialCharTok{\%\textgreater{}\%} \FunctionTok{mutate}\NormalTok{(}\AttributeTok{Day =}
\DecValTok{1}\SpecialCharTok{:}\DecValTok{73}\NormalTok{,}\AttributeTok{centrality=} \StringTok{"inDegree\_Centrality"}\NormalTok{, }\AttributeTok{.before=}\NormalTok{1L)}\OtherTok{{-}\textgreater{}}
\NormalTok{inDegree\_Centrality\_DF}

\NormalTok{OutDegree\_Centrality }\SpecialCharTok{\%\textgreater{}\%} \FunctionTok{as.data.frame}\NormalTok{() }\SpecialCharTok{\%\textgreater{}\%} \FunctionTok{mutate}\NormalTok{(}\AttributeTok{Day =}
\DecValTok{1}\SpecialCharTok{:}\DecValTok{73}\NormalTok{,}\AttributeTok{centrality=} \StringTok{"OutDegree\_Centrality"}\NormalTok{, }\AttributeTok{.before=}\NormalTok{1L)}\OtherTok{{-}\textgreater{}}
\NormalTok{OutDegree\_Centrality\_DF}

\FunctionTok{rbind}\NormalTok{(Degree\_Centrality\_DF, inDegree\_Centrality\_DF,}
\NormalTok{OutDegree\_Centrality\_DF) }\SpecialCharTok{\%\textgreater{}\%} \FunctionTok{as\_tibble}\NormalTok{()}
\end{Highlighting}
\end{Shaded}

\begin{verbatim}
# A tibble: 219 x 47
     Day centrality    `1`   `5`   `6`   `7`  `11`  `13`  `15`  `17`  `19`  `24`
   <int> <chr>       <dbl> <dbl> <dbl> <dbl> <dbl> <dbl> <dbl> <dbl> <dbl> <dbl>
 1     1 Degree_Cen~     4     2     4     4     1     0     2     1     9     1
 2     2 Degree_Cen~     4     2     5     4     2     0     2     1     9     2
 3     3 Degree_Cen~     4     2     5     4     2     0     2     1     9     2
 4     4 Degree_Cen~     5     2     5     4     2     0     2     1     9     2
 5     5 Degree_Cen~     7     2     6     4     2     0     3     2    10     2
 6     6 Degree_Cen~     7     2     6     4     2     0     3     2    10     2
 7     7 Degree_Cen~     7     2     6     4     2     0     3     2     7     2
 8     8 Degree_Cen~     7     2     6     4     3     0     3     2     8     2
 9     9 Degree_Cen~     8     2     6     4     4     0     3     2     8     2
10    10 Degree_Cen~     9     2     6     4     4     0     3     2     9     2
# i 209 more rows
# i 35 more variables: `26` <dbl>, `27` <dbl>, `29` <dbl>, `30` <dbl>,
#   `34` <dbl>, `35` <dbl>, `36` <dbl>, `41` <dbl>, `44` <dbl>, `49` <dbl>,
#   `50` <dbl>, `53` <dbl>, `54` <dbl>, `58` <dbl>, `60` <dbl>, `61` <dbl>,
#   `62` <dbl>, `63` <dbl>, `64` <dbl>, `67` <dbl>, `68` <dbl>, `88` <dbl>,
#   `92` <dbl>, `98` <dbl>, `100` <dbl>, `116` <dbl>, `137` <dbl>, `198` <dbl>,
#   `219` <dbl>, `223` <dbl>, `234` <dbl>, `310` <dbl>, `432` <dbl>, ...
\end{verbatim}

In the same way, closeness, betweenness and eigenvector centralities can
be computed. Please note that we have a new argument here
\texttt{gmode="graph"} which tells \texttt{snafun} that we would like to
compute these centralities considering the network as undirected. You
can also use the \texttt{gmode="digraph"} to compute the measures on a
directed basis. You may need to use the previous code to convert each of
the resulting time series into a data frame and add the day of the
centralities. The \texttt{snafun} can compute other centrality measures
in the same way e.g., information centrality, Bonacich Power Centrality,
Harary Graph Centrality, Bonacich Power Centrality among others.

\begin{Shaded}
\begin{Highlighting}[]
\NormalTok{Closeness\_Centrality }\OtherTok{\textless{}{-}} \FunctionTok{tSnaStats}\NormalTok{(}
\NormalTok{  Active\_Network,}
  \AttributeTok{snafun =} \StringTok{"closeness"}\NormalTok{,}
  \AttributeTok{start =} \DecValTok{1}\NormalTok{,}
  \AttributeTok{end =} \DecValTok{73}\NormalTok{,}
  \AttributeTok{time.interval =} \DecValTok{1}\NormalTok{,}
  \AttributeTok{aggregate.dur =} \DecValTok{1}\NormalTok{,}
  \AttributeTok{gmode=}\StringTok{"graph"}\NormalTok{)}

\NormalTok{Betweenness\_Centrality }\OtherTok{\textless{}{-}} \FunctionTok{tSnaStats}\NormalTok{(}
\NormalTok{  Active\_Network,}
  \AttributeTok{snafun =} \StringTok{"betweenness"}\NormalTok{,}
  \AttributeTok{start =} \DecValTok{1}\NormalTok{,}
  \AttributeTok{end =} \DecValTok{73}\NormalTok{,}
  \AttributeTok{time.interval =} \DecValTok{1}\NormalTok{,}
  \AttributeTok{aggregate.dur =} \DecValTok{1}\NormalTok{,}
  \AttributeTok{gmode=}\StringTok{"graph"}\NormalTok{)}

\NormalTok{Eigen\_Centrality }\OtherTok{\textless{}{-}} \FunctionTok{tSnaStats}\NormalTok{(}
\NormalTok{  Active\_Network,}
  \AttributeTok{snafun =} \StringTok{"evcent"}\NormalTok{,}
  \AttributeTok{start =} \DecValTok{1}\NormalTok{,}
  \AttributeTok{end =} \DecValTok{73}\NormalTok{,}
  \AttributeTok{time.interval =} \DecValTok{1}\NormalTok{,}
  \AttributeTok{aggregate.dur =} \DecValTok{1}\NormalTok{,}
  \AttributeTok{gmode=}\StringTok{"graph"}\NormalTok{)}
\end{Highlighting}
\end{Shaded}

Reachability is another important measure that temporal networks offer.
You can calculate who, when, and after how many steps an interaction
reaches from a certain vertex to another or to every other node in the
network (if it at all does). In the next code, we compute the forward
pathway (earliest reachable node) from node 44 (one of the course
facilitators) using the function \texttt{tPath}. The output is a time
series with all the time distance values and the number of steps.

\begin{Shaded}
\begin{Highlighting}[]
\NormalTok{FwdPathway }\OtherTok{\textless{}{-}} \FunctionTok{tPath}\NormalTok{(}
\NormalTok{  Active\_Network,}
  \AttributeTok{v =} \DecValTok{44}\NormalTok{,}
  \AttributeTok{start =} \DecValTok{0}\NormalTok{,}
  \AttributeTok{graph.step.time =} \DecValTok{7}\NormalTok{,}
  \AttributeTok{end =} \DecValTok{30}\NormalTok{,}
  \AttributeTok{direction =} \StringTok{"fwd"}\NormalTok{)}

\NormalTok{FwdPathway }\SpecialCharTok{\%\textgreater{}\%} \FunctionTok{as\_tibble}\NormalTok{()}
\end{Highlighting}
\end{Shaded}

\begin{verbatim}
# A tibble: 45 x 7
    tdist previous gsteps start   end direction type           
    <dbl>    <dbl>  <dbl> <dbl> <dbl> <chr>     <chr>          
 1   7.09       44      1     0    30 fwd       earliest.arrive
 2  32.9        45      4     0    30 fwd       earliest.arrive
 3  34.8        31      3     0    30 fwd       earliest.arrive
 4  14.1         1      2     0    30 fwd       earliest.arrive
 5  16.1        44      1     0    30 fwd       earliest.arrive
 6  30.9        13      2     0    30 fwd       earliest.arrive
 7  28.1        23      4     0    30 fwd       earliest.arrive
 8 Inf           0    Inf     0    30 fwd       earliest.arrive
 9   7.28       44      1     0    30 fwd       earliest.arrive
10  24.2         9      2     0    30 fwd       earliest.arrive
# i 35 more rows
\end{verbatim}

\begin{figure}

{\centering \includegraphics[width=3.33333in,height=\textheight]{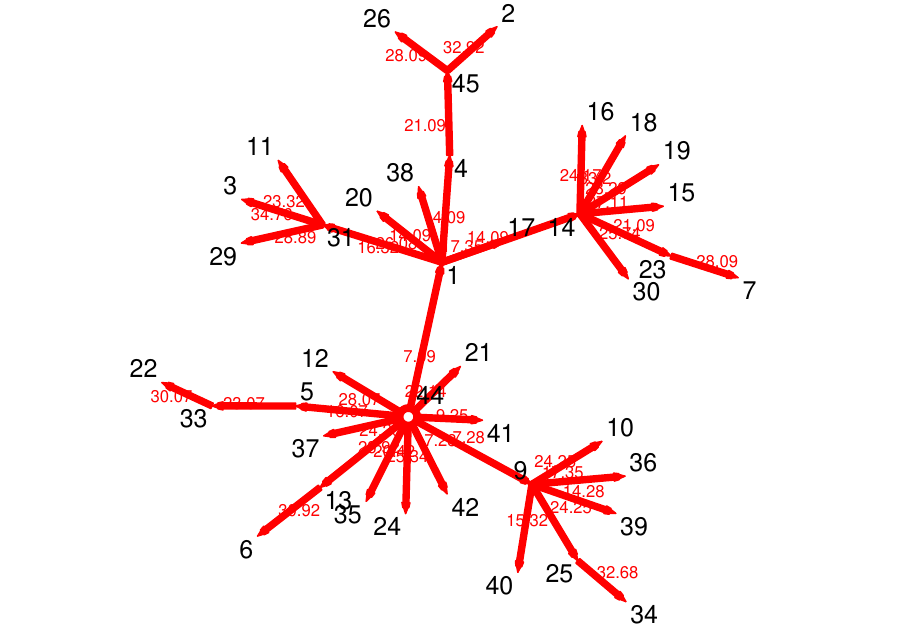}

}

\caption{\label{fig-fwdpathway}A forward pathway (earliest reachable
node) pathway of interactions with node 44}

\end{figure}

More importantly, we can plot the hierarchical path from the given node,
by simply using \texttt{plot} function, see Figure~\ref{fig-fwdpathway}.

\begin{Shaded}
\begin{Highlighting}[]
\FunctionTok{plot}\NormalTok{(FwdPathway)}
\end{Highlighting}
\end{Shaded}

\begin{figure}[H]

{\centering \includegraphics{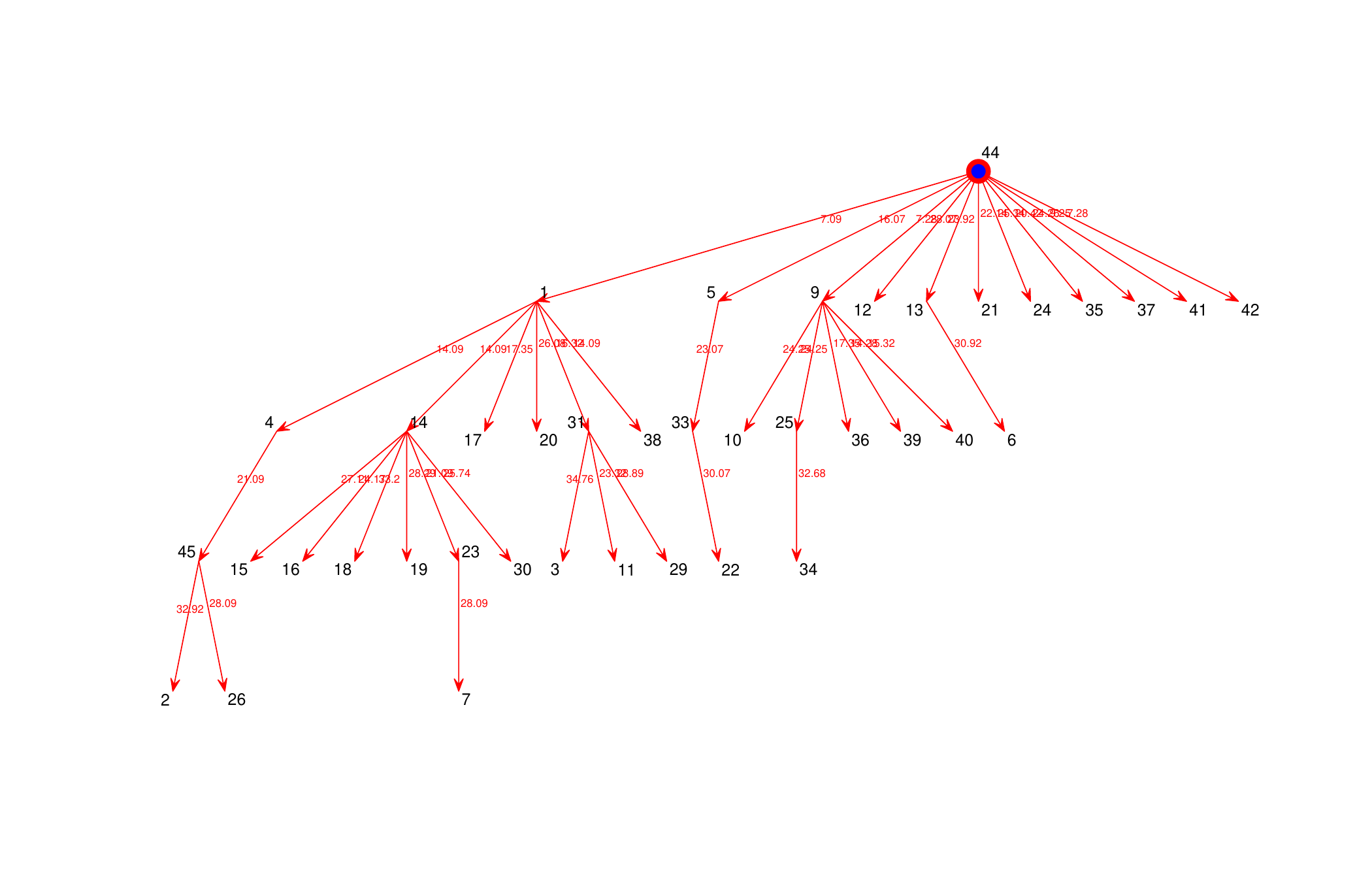}

}

\caption{\label{fig-fwdpathway-gvizplotPaths}An improved forward pathway
with Graphviz}

\end{figure}

We may also use \texttt{Graphviz} to make the plot look better and
hierarchical (Figure~\ref{fig-fwdpathway-gvizplotPaths}).

\begin{Shaded}
\begin{Highlighting}[]
\FunctionTok{plot}\NormalTok{(FwdPathway, }\AttributeTok{edge.lwd =} \FloatTok{0.1}\NormalTok{, }\AttributeTok{vertex.col=} \StringTok{"blue"}\NormalTok{, }\AttributeTok{pad =} \SpecialCharTok{{-}}\DecValTok{4}\NormalTok{,}
     \AttributeTok{coord=}\FunctionTok{network.layout.animate.Graphviz}\NormalTok{(}\FunctionTok{as.network}\NormalTok{(FwdPathway), }
                   \AttributeTok{layout.par =} \FunctionTok{list}\NormalTok{(}\AttributeTok{gv.engine=}\StringTok{\textquotesingle{}dot\textquotesingle{}}\NormalTok{,}
                              \AttributeTok{gv.args=}\StringTok{\textquotesingle{}{-}Granksep=2\textquotesingle{}}\NormalTok{)))}
\end{Highlighting}
\end{Shaded}

\begin{figure}[H]

{\centering \includegraphics[width=4in,height=\textheight]{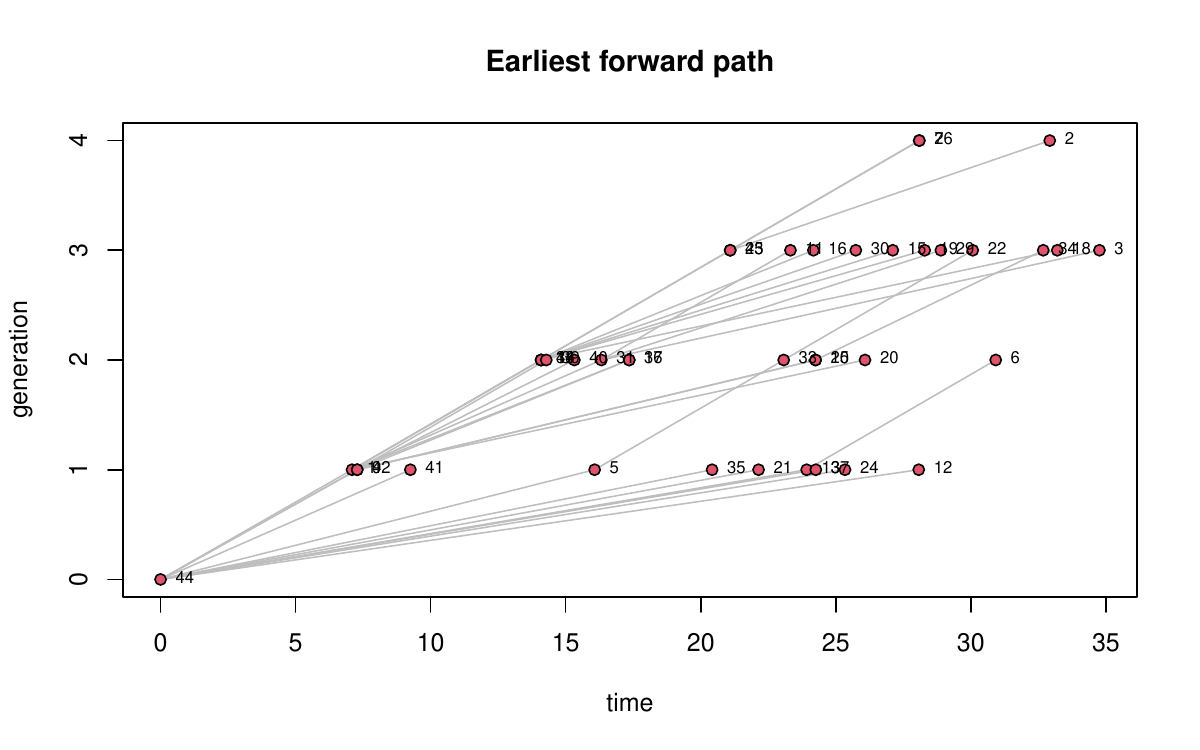}

}

\caption{\label{fig-transmission-timeline}Transmission hierarchical tree
showing the earliest forward interaction with node 44}

\end{figure}

Another option is to plot the network diffusion or transmission
hierarchical tree with generation time vs.~clock/model time using
\texttt{transmissionTimeline} function as in
Figure~\ref{fig-transmission-timeline}. For more on these functions,
readers are invited to read the function manuals and for usage, these
papers offer a starting point {[}3, 9{]}.

\begin{Shaded}
\begin{Highlighting}[]
\FunctionTok{transmissionTimeline}\NormalTok{(FwdPathway, }\AttributeTok{jitter =}\NormalTok{ F, }\AttributeTok{displaylabels =} \ConstantTok{TRUE}\NormalTok{, }
                    \AttributeTok{main =} \StringTok{"Earliest forward path"}\NormalTok{ )}
\end{Highlighting}
\end{Shaded}

\hypertarget{discussion}{%
\section{Discussion}\label{discussion}}

Learning can be viewed as relational, interdependent, and temporal and
therefore, methods that account for such multifaceted dynamic processes
are required {[}1, 3{]}. We have shown the main advantages of temporal
networks and the potentials it offers for modeling dynamic learning
processes. These potentials or features can facilitate the modeling of
the complex natural processes--including the emergence, evolution,
diffusion or disappearance of learners' activities, communities or
social processes that unfold over time. Such features can augment the
existing analytics method and help shed light on many learning phenomena
{[}16{]}. Taking advantage of time dynamics allows us temporal evolution
of co-construction of knowledge, the flow of information and the
building of social relationships, to name a few examples. What is more,
temporal networks allow the longitudinal modeling and analysis of
interactions across longer periods of time e.g., full duration of a
course, project or meeting using time-respecting paths {[}9, 15{]}.

There are several methods that can harness the temporal dimensions of a
learning process, e.g., process and sequence mining, time series methods
and epistemic network analysis {[}27, 28{]}. While such methods have
given a wealth of information and insights about learning processes,
they fall short when it comes to the relational aspects {[}9{]}. We
review here and in short the main differences between such methods of
temporal networks. Process mining is a method for the discovery and
modeling of a temporal process {[}1, 29{]}.Yet, the relational aspect is
completely ignored. The case is similar for sequence mining where the
time-ordered sequences are modeled regardless of the theri interactions
{[}27, 30, 31{]}. Epistemic network analysis is another method that
allows the study of co-temporal interactions. However, the ``temporal
aspect'' is limited to combining data within a temporal window and later
modeling the interactions as a static network. Put another way,
Epistemic network analysis is a special type of static networks where
edges are defined based on co-occurrence {[}32{]}. For a comparison
between the various methods see {[}3{]}.

\hypertarget{learning-resources}{%
\section{Learning resources}\label{learning-resources}}

A good place to start is to get acquainted with the cited research in
the literature review section. Other good references could be the
methodological paper that gives a detailed overview of temporal networks
which some parts of this chapter has been built around it {[}33{]}.
There are few, yet very informative tutorials that we can suggest, most
notable are the tutorials by {[}34{]} and {[}18{]}. The packages used in
this chapter have very informative manuals: TSNA{[}19{]}, NDTV{[}20{]}
and networkDynamic{[}35{]}. Some seminal papers can be recommended here,
especially the following papers and books.

\begin{itemize}
\item
  Holme, P. (2015). Modern temporal network theory: a colloquium.
  European Physical Journal B, 88(9).
\item
  Holme, P., \& Saramäki, J. (2019). A Map of Approaches to Temporal
  Networks (pp.~1--24).
\item
  Holme, P., \& Saramäki, J. (Eds.). (2019). Temporal network theory
  (Vol. 2). New York: Springer.
\end{itemize}

\hypertarget{references}{%
\section{References}\label{references}}

\hypertarget{refs}{}
\begin{CSLReferences}{0}{0}
\leavevmode\vadjust pre{\hypertarget{ref-Saqr_Lopez-Pernas_2023}{}}%
\CSLLeftMargin{1. }%
\CSLRightInline{Saqr M, Lopez-Pernas S (2023) The temporal dynamics of
online problem-based learning: Why and when sequence matters.
International Journal of Computer-Supported Collaborative Learning
18:11--37}

\leavevmode\vadjust pre{\hypertarget{ref-Johnson_Azevedo_DMello_2011}{}}%
\CSLLeftMargin{2. }%
\CSLRightInline{Johnson AM, Azevedo R, D'Mello SK (2011) The temporal
and dynamic nature of self-regulatory processes during independent and
externally assisted hypermedia learning. Cognition and instruction
29:471--504}

\leavevmode\vadjust pre{\hypertarget{ref-Saqr_Peeters_Viberg_2021}{}}%
\CSLLeftMargin{3. }%
\CSLRightInline{Saqr M, Peeters W, Viberg O (2021) The relational,
co-temporal, contemporaneous, and longitudinal dynamics of
self-regulation for academic writing. Research and Practice in
Technology Enhanced Learning 16:29}

\leavevmode\vadjust pre{\hypertarget{ref-Chen_Wise_Knight_Cheng_2016}{}}%
\CSLLeftMargin{4. }%
\CSLRightInline{Chen B, Wise AF, Knight S, Cheng BH (2016) Putting
temporal analytics into practice: The 5th international workshop on
temporality in learning data. In: Proceedings of the sixth international
conference on learning analytics \& knowledge. ACM, pp 488--489}

\leavevmode\vadjust pre{\hypertarget{ref-Reimann_2009}{}}%
\CSLLeftMargin{5. }%
\CSLRightInline{Reimann P (2009) Time is precious: Variable- and
event-centred approaches to process analysis in CSCL research.
International Journal of Computer-Supported Collaborative Learning
4:239--257}

\leavevmode\vadjust pre{\hypertarget{ref-Chen_Poquet_2020}{}}%
\CSLLeftMargin{6. }%
\CSLRightInline{Chen B, Poquet O (2020) Socio-temporal dynamics in peer
interaction events. ACM International Conference Proceeding Series
203--208}

\leavevmode\vadjust pre{\hypertarget{ref-Lopez-Pernas_Saqr_Gordillo_Barra_2022}{}}%
\CSLLeftMargin{7. }%
\CSLRightInline{Lopez-Pernas S, Saqr M, Gordillo A, Barra E (2022) A
learning analytics perspective on educational escape rooms. Interactive
Learning Environments 1--17}

\leavevmode\vadjust pre{\hypertarget{ref-Saqr_Lopez-Pernas_2022d}{}}%
\CSLLeftMargin{8. }%
\CSLRightInline{Saqr M, Lopez-Pernas S (2022) How CSCL roles emerge,
persist, transition, and evolve over time: A four-year longitudinal
study. Computers \& education 189:104581}

\leavevmode\vadjust pre{\hypertarget{ref-Saqr_Peeters_2022}{}}%
\CSLLeftMargin{9. }%
\CSLRightInline{Saqr M, Peeters W (2022) Temporal networks in
collaborative learning: A case study. British Journal of Educational
Technology. \url{https://doi.org/10.1111/bjet.13187}}

\leavevmode\vadjust pre{\hypertarget{ref-Holme_2015}{}}%
\CSLLeftMargin{10. }%
\CSLRightInline{Holme P (2015) Modern temporal network theory: A
colloquium. European Physical Journal B: Condensed Matter and Complex
Systems 88:
https://doi.org/\href{https://doi.org/10.1140/epjb/e2015-60657-4}{10.1140/epjb/e2015-60657-4}}

\leavevmode\vadjust pre{\hypertarget{ref-Holme_Saramaki_2012}{}}%
\CSLLeftMargin{11. }%
\CSLRightInline{Holme P, Saramäki J (2012) Temporal networks. Physics
reports 519:97--125}

\leavevmode\vadjust pre{\hypertarget{ref-Holme_Saramuxe4ki_2019}{}}%
\CSLLeftMargin{12. }%
\CSLRightInline{Holme P, Saramäki J (2019) A map of approaches to
temporal networks. pp 1--24}

\leavevmode\vadjust pre{\hypertarget{ref-Nicosia_Tang_Mascolo_Musolesi_Russo_Latora_2013}{}}%
\CSLLeftMargin{13. }%
\CSLRightInline{Nicosia V, Tang J, Mascolo C, Musolesi M, Russo G,
Latora V (2013) Graph metrics for temporal networks. In: Temporal
networks. Springer, pp 15--40}

\leavevmode\vadjust pre{\hypertarget{ref-Li_Cornelius_Liu_Wang_Barabuxe1si_2017}{}}%
\CSLLeftMargin{14. }%
\CSLRightInline{Li A, Cornelius SP, Liu Y-Y, Wang L, Barabási A-L (2017)
The fundamental advantages of temporal networks. Science 358:1042--1046}

\leavevmode\vadjust pre{\hypertarget{ref-Saqr_Nouri_2020}{}}%
\CSLLeftMargin{15. }%
\CSLRightInline{Saqr M, Nouri J (2020) High resolution temporal network
analysis to understand and improve collaborative learning. In:
Proceedings of the tenth international conference on learning analytics
\& knowledge. ACM, New York, NY, USA, pp 314--319}

\leavevmode\vadjust pre{\hypertarget{ref-Saqr_Lopez-Pernas_2022a}{}}%
\CSLLeftMargin{16. }%
\CSLRightInline{Saqr M, Lopez-Pernas S (2022) Instant or distant: A
temporal network tale of two interaction platforms and their influence
on collaboration. In: Educating for a new future: Making sense of
technology-enhanced learning adoption. Springer International
Publishing, pp 594--600}

\leavevmode\vadjust pre{\hypertarget{ref-Saqr_Angel2024}{}}%
\CSLLeftMargin{17. }%
\CSLRightInline{Saqr M, Lopez-Pernas S, Conde MÁ, Hernández-García Á
(2024) Social network analysis: A primer, a guide and a tutorial in r.
In: Saqr M, Lopez-Pernas S (eds) Learning analytics methods and
tutorials: A practical guide using r. Springer, pp in--press}

\leavevmode\vadjust pre{\hypertarget{ref-Bender-deMoll_2016}{}}%
\CSLLeftMargin{18. }%
\CSLRightInline{Bender-deMoll S (2016)
\href{https://web.archive.org/web/20180423112846/http://statnet.csde.washington.edu/workshops/SUNBELT/current/ndtv/ndtv_workshop.html}{Temporal
network tools in statnet: networkDynamic, ndtv and tsna}. statnet}

\leavevmode\vadjust pre{\hypertarget{ref-Bender-deMoll_Morris_2016}{}}%
\CSLLeftMargin{19. }%
\CSLRightInline{Bender-deMoll S, Morris M (2016) Tsna: Tools for
temporal social network analysis. R package version 02 0, URL
https://CRAN R-project org/package= tsna}

\leavevmode\vadjust pre{\hypertarget{ref-Bender-deMoll_2018}{}}%
\CSLLeftMargin{20. }%
\CSLRightInline{Bender-deMoll S (2018)
\href{https://cran.r-project.org/package=ndtv}{Ndtv: Network dynamic
temporal visualizations}}

\leavevmode\vadjust pre{\hypertarget{ref-Lopez-Pernas_Saqr_Del}{}}%
\CSLLeftMargin{21. }%
\CSLRightInline{Lopez-Pernas S, Saqr M, Del Rio L (2024) A broad
collection of datasets for educational research training and
application. In: Saqr M, Lopez-Pernas S (eds) Learning analytics methods
and tutorials: A practical guide using r. Springer, pp in--press}

\leavevmode\vadjust pre{\hypertarget{ref-Vu_Pattison_Robins_2015}{}}%
\CSLLeftMargin{22. }%
\CSLRightInline{Vu D, Pattison P, Robins G (2015) Relational event
models for social learning in MOOCs. Social networks 43:121--135}

\leavevmode\vadjust pre{\hypertarget{ref-Hernuxe1ndez-Garcuxeda_Gonzuxe1lez-Gonzuxe1lez_Jimuxe9nez-Zarco_Chaparro-Peluxe1ez_2015}{}}%
\CSLLeftMargin{23. }%
\CSLRightInline{Hernández-García Á, González-González I, Jiménez-Zarco
AI, Chaparro-Peláez J (2015) Applying social learning analytics to
message boards in online distance learning: A case study. Computers in
human behavior 47:68--80}

\leavevmode\vadjust pre{\hypertarget{ref-Poquet_Saqr_Chen_2021}{}}%
\CSLLeftMargin{24. }%
\CSLRightInline{Poquet O, Saqr M, Chen B (2021) Recommendations for
network research in learning analytics: To open a conversation. In:
Proceedings of the NetSciLA21 workshop}

\leavevmode\vadjust pre{\hypertarget{ref-Saqr_Elmoazen_Tedre_Lopez-Pernas_Hirsto_2022}{}}%
\CSLLeftMargin{25. }%
\CSLRightInline{Saqr M, Elmoazen R, Tedre M, Lopez-Pernas S, Hirsto L
(2022) How well centrality measures capture student achievement in
computer-supported collaborative learning? -- a systematic review and
meta-analysis. Educational Research Review 35:100437}

\leavevmode\vadjust pre{\hypertarget{ref-Saqr_Lopez-Pernas_2022c}{}}%
\CSLLeftMargin{26. }%
\CSLRightInline{Saqr M, Lopez-Pernas S (2022) The curious case of
centrality measures: A large-scale empirical investigation. Journal of
learning analytics 9:13--31}

\leavevmode\vadjust pre{\hypertarget{ref-Lopez-Pernas_Saqr_2021}{}}%
\CSLLeftMargin{27. }%
\CSLRightInline{Lopez-Pernas S, Saqr M (2021) Bringing synchrony and
clarity to complex multi-channel data: A learning analytics study in
programming education. IEEE Access 1--1}

\leavevmode\vadjust pre{\hypertarget{ref-Peeters_Saqr_Viberg_2020}{}}%
\CSLLeftMargin{28. }%
\CSLRightInline{Peeters W, Saqr M, Viberg O (2020) Applying learning
analytics to map students ' self-regulated learning tactics in an
academic writing course}

\leavevmode\vadjust pre{\hypertarget{ref-Vartiainen_Lopez-Pernas_Saqr_Kahila_Parkki_Tedre_Valtonen_2022}{}}%
\CSLLeftMargin{29. }%
\CSLRightInline{Vartiainen H, Lopez-Pernas S, Saqr M, Kahila J, Parkki
T, Tedre M, Valtonen T (2022) Mapping students' temporal pathways in a
computational thinking escape room. Proceedings http://ceur-ws org ISSN
1613:0073}

\leavevmode\vadjust pre{\hypertarget{ref-Heikkinen_Lopez-Pernas_Malmberg_Tedre_Saqr}{}}%
\CSLLeftMargin{30. }%
\CSLRightInline{Heikkinen S, Lopez-Pernas S, Malmberg J, Tedre M, Saqr M
\href{https://ceur-ws.org/Vol-3383/FLAIEC22_paper_2583.pdf}{How do
business students self-regulate their project management learning? A
sequence mining study}}

\leavevmode\vadjust pre{\hypertarget{ref-Saqr_Lopez-Pernas_Jovanoviux107_Gaux161eviux107_2023}{}}%
\CSLLeftMargin{31. }%
\CSLRightInline{Saqr M, Lopez-Pernas S, Jovanović J, Gašević D (2023)
Intense, turbulent, or wallowing in the mire: A longitudinal study of
cross-course online tactics, strategies, and trajectories. The Internet
and Higher Education 57:100902}

\leavevmode\vadjust pre{\hypertarget{ref-Elmoazen_Saqr_Tedre_Hirsto_2022}{}}%
\CSLLeftMargin{32. }%
\CSLRightInline{Elmoazen R, Saqr M, Tedre M, Hirsto L (2022) A
systematic literature review of empirical research on epistemic network
analysis in education. IEEE access: practical innovations, open
solutions 10:17330--17348}

\leavevmode\vadjust pre{\hypertarget{ref-Saqr_Lopez-Pernas_2022b}{}}%
\CSLLeftMargin{33. }%
\CSLRightInline{Saqr M, Lopez-Pernas S (2022)
\href{https://www.researchgate.net/profile/Mohammed-Saqr/publication/364997941_The_Why_the_What_and_the_How_to_Model_a_Dynamic_Relational_Learning_Process_with_Temporal_Networks/links/636271222f4bca7fd0270b74/The-Why-the-What-and-the-How-to-Model-a-Dynamic-Relational-Learning-Process-with-Temporal-Networks.pdf}{The
why, the what and the how to model a dynamic relational learning process
with temporal networks}. In: Proceedings of the NetSciLA22 workshop}

\leavevmode\vadjust pre{\hypertarget{ref-Brey_2018}{}}%
\CSLLeftMargin{34. }%
\CSLRightInline{Brey A (2018) Temporal network analysis with r. The
programming historian. \url{https://doi.org/10.46430/phen0080}}

\leavevmode\vadjust pre{\hypertarget{ref-Butts_Leslie-Cook_Krivitsky_Bender-Demoll_2014}{}}%
\CSLLeftMargin{35. }%
\CSLRightInline{Butts CT, Leslie-Cook A, Krivitsky PN, Bender-Demoll S
(2014) Dynamic extensions for network objects}

\end{CSLReferences}

\end{document}